\documentclass[]{article}

\makeatletter
\newcommand{\publicationtitle}{How Far Does \Los's Theorem Extend To Kripke--Joyal Semantics?}%
\newcommand{\publicationsubtitle}{Sufficient Conditions, Counterexamples, and a Conjecture}%
\newcommand{\publicationabstract}{%
\Los's theorem, also known as the fundamental result of ultraproducts, states that the ultraproduct over a family of structures for the same language satisfies a first-order formula if and only if the set of indices for which the structures satisfy the formula belongs to the underlying ultrafilter. The associated notion of satisfaction is the Tarskian one via the elements of the set-theoretic structure that allow interpreting the formula. In the context of topoi, Kripke--Joyal semantics extends Tarski's notion to categorical logic. In this article, we investigate \Los's theorem for first-order structures on locally presentable topoi with Kripke--Joyal semantics. More precisely, we identify a set of structural properties on the ambient topos (two-valuedness and projectivity of the terminal object) used to derive a categorical analog of \Los's theorem. As an application, we derive compactness results for several fragments of
first-order logic. We also show via explicit counterexamples that our structural assumptions are only sufficient. The counterexamples suggest an external approach and we conjecture that \Los's property for full first-order logic is equivalent to the existence of a conservative family of logical points compatible with the ultraproduct construction.
}
\makeatother

\usepackage{
    amsmath,
    amssymb,
    amsthm,
    MnSymbol, %
    amsfonts,
}

\usepackage[inline]{enumitem}

\newtheorem{Definition}{Definition}[section]
\newtheorem{Example}[Definition]{Example}

\newtheorem{Theorem}[Definition]{Theorem}
\newtheorem{Proposition}[Definition]{Proposition}
\newtheorem{Lemma}[Definition]{Lemma}
\newtheorem{Corollary}[Definition]{Corollary}
\newtheorem{Conjecture}[Definition]{Conjecture}
\newtheorem{Notation}[Definition]{Notation}

\usepackage[utf8]{inputenc}
\usepackage{csquotes}
\usepackage{color}

\usepackage{quiver}

\usepackage{url}
\usepackage[
    pdftex,
    pdfborder={0 0 0},
]{hyperref}

\usepackage{cleveref}
\Crefname{section}{Sect.}{Sects.}
\crefname{section}{Sect.}{Sects.}
\Crefname{Definition}{Def.}{Defs.}
\crefname{Definition}{Def.}{Defs.}
\Crefname{Proposition}{Prop.}{Props.}
\crefname{Proposition}{Prop.}{Props.}
\Crefname{Example}{Ex.}{Exs.}
\crefname{Example}{Ex.}{Exs.}
\Crefname{Theorem}{Thm.}{Thms.}
\crefname{Theorem}{Thm.}{Thms.}
\Crefname{Corollary}{Cor.}{Cors.}
\crefname{Corollary}{Cor.}{Cors.}

\usepackage[
    backend=biber,
    style=numeric,
    sorting=nyt,
    date=year,
    giveninits=true,
    maxbibnames=99,
]{biblatex}
\bibliography{biblio}

\usepackage{orcidlink}

\newcommand{\subjclass}[2][2000]{%
\textit{#1 Mathematics Subject Classification.} #2
}
\newcommand{\keywords}[1]{%
\textit{Keywords.} #1
}

\newcommand{\from}{\:{\colon}\linebreak[0]\:}									%

\makeatletter
\def\bstr{b}
\def\bfstr{bf}
\def\cstr{c}
\def\fstr{f}
\def\sstr{s}

\def\strLst{A,B,C,D,d,E,F,G,H,I,J,K,L,M,N,O,P,Q,R,S,T,U,V,W,X,Y,Z}

\newcommand{\MkB}[1]{\expandafter\def\csname\bstr#1\endcsname{\mathbb{#1}}}
  \@for\i:=\strLst\do{%
    \expandafter\MkB \i     }
    
\newcommand{\MkBF}[1]{\expandafter\def\csname\bfstr#1\endcsname{\mathbf{#1}}}
  \@for\i:=\strLst\do{%
    \expandafter\MkBF \i     }

\newcommand{\MkCal}[1]{\expandafter\def\csname\cstr#1\endcsname{\mathcal{#1}}}
  \@for\i:=\strLst\do{%
    \expandafter\MkCal \i     }
    
\newcommand{\MkFrak}[1]{\expandafter\def\csname\fstr#1\endcsname{\mathfrak{#1}}}
  \@for\i:=\strLst\do{%
    \expandafter\MkFrak \i     }
    
\newcommand{\MkSF}[1]{\expandafter\def\csname\sstr#1\endcsname{\mathsf{#1}}}
  \@for\i:=\strLst\do{%
    \expandafter\MkSF \i     }
    
\makeatother

\newcommand{\Functor}[8][cccc]
{
	\ensuremath{
		\ifthenelse{\equal{#2}{}}{}{#2 : \left\{ }
		\begin{array}{#1}
			#3 	&	\longrightarrow 	& 	#4 		\\
			#5 	& 	\longmapsto 		& 	#6		\\
			#7	&	\longmapsto			&	#8
		\end{array}
		\ifthenelse{\equal{#2}{}}{}{\right.}
	}
}

\newcommand{\Function}[6][\longrightarrow]
{%
	\ensuremath{%
		\ifthenelse{\equal{#2}{}}{}{#2 : \left\lbrace }%
		\begin{array}{ccc}%
			#3 	&	 #1	& 	#4 		\\%
			#5 	& 	\longmapsto 		& 	#6%
		\end{array}%
		\ifthenelse{\equal{#2}{}}{}{\right.}%
	}%
}

\newcommand{\Set}{\mathrm{Set}}                                             %
\newcommand{\set}[1]{\left\{#1\right\}}                                     %

\newcommand{\Sub}{\operatorname{\mathrm{Sub}}}
\newcommand{\ob}[1]{\operatorname{\mathrm{Ob}}(#1)}  						%
\newcommand{\Hom}{\operatorname{\mathrm{Hom}}}                              %
\newcommand{\id}{\mathrm{id}}                                               %

\newcommand{\dom}{\operatorname{\mathrm{dom}}}                              %
\newcommand{\codom}{\operatorname{\mathrm{cod}}}                            %
\newcommand{\im}{\mathrm{Im}}												%
\newcommand{\colim}{\mathrm{colim}}											%

\newcommand\Omit[1]{}

\newcommand{\sem}[2][\cM]{[\![ #2 ]\!]_{#1}}

\mathchardef\mhyphen="2D 													%

\newcommand{\Pos}{\mathrm{Pos}}
\newcommand{\Los}{\L{}o{\'s}}
\newcommand{\mto}{\rightarrowtail}

\newcommand{\initial}{\mathbf{\varnothing}}						            %
\newcommand{\terminal}{\mathfrak{1}}										%
\newcommand{\Structures}[2][\Sigma]{#1\mhyphen\linebreak[0]\mathrm{Str}(#2)}                  %
\newcommand{\SatisIndexSet}[2][\alpha]{\mathrm{SatInd}_{#1}(#2)}

\newcommand{\emptycontext}{[\,]}
\newcommand*\mathinhead[2]{\texorpdfstring{\(\boldsymbol{#1}\)}{#2}} \newif\ifcomment
\commenttrue
\ifcomment

\usepackage[%
    draft,
    colorinlistoftodos,
    backgroundcolor=white,
    textsize=footnotesize
    ]{todonotes}

\makeatletter
 \providecommand\@dotsep{5}
 \def\listtodoname{List of Todos}
 \def\listoftodos{\@starttoc{tdo}\listtodoname}
 \makeatother

\makeatletter
\def\ifemptyarg#1{%
  \if\relax\detokenize{#1}\relax %
    \expandafter\@firstoftwo
  \else
    \expandafter\@secondoftwo
  \fi}
\makeatother

\newcounter{todoListItemS}

\def\commStr{Comm}
\def\todoStr{Todo}

\newcommand{\defineCommMacro}[2]{%
  \expandafter\newcommand\csname #1\commStr\endcsname[2][]{%
\ifemptyarg{##1}{%
  \vspace{0.5em}
  \stepcounter{todoListItemS}
        \todo[inline,bordercolor=#2]{%
        \textbf{{\color{#2}Comment~[\uppercase{#1}~\thetodoListItemS]:}}~##2%
        }
  \vspace{0.5em
  }}{%
  \vspace{0.5em}
    \stepcounter{todoListItemS}
        \todo[%
        		inline,
		bordercolor=#2, 
		caption={\textbf{{\color{#2}Comment~[\uppercase{#1}~\thetodoListItemS]:~##1}}}
	]{%
        \textbf{{\color{#2}Comment~[\uppercase{#1}~\thetodoListItemS]:}}~##2%
        }
  \vspace{0.5em}
        }
        }}
    
\newcommand{\defineTodoMacro}[2]{%
  \expandafter\newcommand\csname #1\todoStr\endcsname[2][]{%
\ifemptyarg{##1}{%
\stepcounter{todoListItemS}
  \vspace{0.5em}
        \todo[inline,bordercolor=#2]{%
        \textbf{{\color{#2}Todo~[\uppercase{#1}~\thetodoListItemS]:}}~##2%
        }
  \vspace{0.5em
  }}{%
  \stepcounter{todoListItemS}
  \vspace{0.5em}
        \todo[inline,bordercolor=#2, caption={\textbf{{\color{#2}Todo~[\uppercase{#1}~\thetodoListItemS]:~##1}}}]{%
        \textbf{{\color{#2}Todo~[\uppercase{#1}~\thetodoListItemS]:}}~##2%
        }
  \vspace{0.5em}
        }
        } 
}

\usepackage[deletedmarkup=sout]{changes}

\newcommand{\changesAuthorSetup}[3]{%
\definechangesauthor[name={#1}, color=#3]{#2}
\expandafter\newcommand\csname#2A\endcsname[1]{%
    \added[id=#2]{##1}
}
\expandafter\newcommand\csname#2D\endcsname[1]{%
    \added[id=#2]{##1}
}
\expandafter\newcommand\csname#2R\endcsname[2]{%
    \added[id=#2]{##1}{##2}
}
}

\colorlet{MAcolor}{red!70!black}  %

\defineCommMacro{ma}{MAcolor}
\defineTodoMacro{ma}{MAcolor}

\changesAuthorSetup{M. Aiguier}{ma}{MAcolor}

\colorlet{RPcolor}{green!70!black}  %

\defineCommMacro{rp}{RPcolor}
\defineTodoMacro{rp}{RPcolor}

\changesAuthorSetup{R. Pascual}{rp}{RPcolor}

\newcommand{\registeruser}[2]{%
  \expandafter\newcommand\csname #1\endcsname[1]{\textcolor{#2}{##1}}
  \newenvironment{#1-env}{\color{#2}}{}
}

\usepackage[normalem]{ulem}
\usepackage{color}
\registeruser{MA}{MAcolor}
\registeruser{RP}{RPcolor}

 \fi

\begin{document}

\title{%
\publicationtitle\\[0.3em]
\Large\publicationsubtitle
}

\author{%
Marc Aiguier$^1$\,\orcidlink{0000-0003-0154-0909}
and Romain Pascual$^{1,2}$\,\orcidlink{0000-0003-1282-1933}\\
1. MICS, CentraleSupelec, Universit\'e Paris Saclay, France \\
2. Karlsruhe Institute of Technology, Karlsruhe, Germany \\
{\it marc.aiguier@centralesupelec.fr}\\
{\it romain.pascual@centralesupelec.fr}
}

\date{}

\maketitle

\begin{abstract}
\publicationabstract

\medskip
\noindent
\subjclass[2020]{%
03G30, %
03C20, %
03C95, %
18B25, %
}

\smallskip
\noindent
\keywords{%
Topoi ; 
Categorical logic ; 
Ultraproducts ; 
\Los's theorem ; 
Kripke--Joyal semantics.
}
\end{abstract}

\tableofcontents

\section{Introduction}
\label{sec:introduction}

\Los's theorem is a result in model theory, also known as the fundamental result of ultraproducts. In its standard form, \Los's theorem states that the ultraproduct of a family of structures for the same language satisfies a first-order formula if and only if the set of indices for which the structures satisfy the formula belongs to the underlying ultrafilter. In this article, we study how to extend this result to first-order structures interpreted in elementary topoi for Kripke--Joyal semantics. We also show that the extension entails its set-theoretic version. Finally, we discuss the connection with the compactness theorem, and, more importantly, the precise limits of our sufficient conditions which we formulate as an open problem.

In its set-based instance, \Los's theorem can be formulated as follows:
\begin{Theorem}[\Los's Theorem~\cite{los_quelques_1955}]
\label{thm:Los:set-theory}
Let \((\cM_i)_{i\in I}\) be an \(I\)-indexed family of nonempty \(\Sigma\)-structures, and let \(F\) be an ultrafilter on \(I\). Let \(\prod_F \cM\) be the ultraproduct of \((\cM_i)_{i\in I}\) with respect to \(F\). Since each \(\cM_i\) is nonempty, the ultraproduct \(\prod_F \cM\) is the quotient of \(\prod_{i \in I} \cM_i\) by the equivalence relation \(\sim_F\) identifying \(I\)-sequences that coincide on a set of indices\footnote{That is $(a_i)_I \sim_F (b_i)_{i \in I}$ if and only if $\set{i \in I \mid a_i = b_i} \in F$} belonging to \(F\). Let \((a^k_i)_{i \in I}\) be \(I\)-sequences for \(k \in \set{1, \ldots, n}\), with \([a^k]_{\sim_F}\) denoting their equivalence classes. Then for each \(\Sigma\)-formula \(\varphi\),
\[
    \prod_F \cM \models \varphi([a^1]_{\sim_F}, \ldots [a^n]_{\sim_F}) ~\mbox{iff}~ \set{j \in I \mid \cM_j \models \varphi(a^1_j, \ldots a^n_j)} \in F.
\]
\end{Theorem}

For instance, consider a family of structures \((\cM_i)_{i\in \bN}\) where each \(\cM_i\) is a model of Peano's arithmetic. If \(F\) is a non-principal ultrafilter on \(\bN\), the ultraproduct \(\prod_F \cM\) is a model of Peano's arithmetic that provides a non-standard model of arithmetic containing ``infinite'' natural numbers. More applications to model theory, algebra, and non-standard analysis are discussed in the surveys by J. Keisler's ~\cite{keisler_ultraproduct_2010} and S. Galbor~\cite{sagi_ultraproducts_2023}. In logic, and more precisely in model theory, the standard corollary of \Los's theorem is the compactness theorem, stating that a set of first-order sentences admits a model if and only if every finite subset of it does.

The theorem considers a first-order signature \(\Sigma = (S, F, R)\), i.e., consisting of a set of sorts, function symbols, and relation symbols, to be used to build \(\Sigma\)-structures. In set theory, these \(\Sigma\)-structures consist of sets associated with functions and relations characterized by the signature. Categorical logic~\cite{Johnstone02,MM12} uses tools from category theory to extend the set-theoretic semantics of first-order logic (FOL) to categories and, more specifically, to a family of categories known as elementary topoi or to specific fragments thereof. A first difficulty occurs: in this framework, we can no longer talk about the elements of \(\Sigma\)-structure carriers as they are not sets anymore.
In particular, \(I\)-sequences do not exist. Then, how can the ultraproduct \(\prod_F \cM\) be defined categorically?
In the category of sets and functions, ultraproducts correspond to filtered products where the underlying filter is an ultrafilter. %
Thus, filtered products correspond to particular instances of the categorical concept of reduced products~\cite{AN78,Dia08,Dia17,Okh66}, i.e., to colimits of directed diagrams of projections between the (direct) products determined by the corresponding filter.

The statement made in the theorem relies on the notion of satisfaction of a formula by a \(\Sigma\)-structure, which unveils a second difficulty: we need a categorical counterpart to this notion of satisfaction.
In the set-theoretic framework, the usual Tarskian notion of satisfaction for a formula \(\varphi\) over \(\Sigma\) in a \(\Sigma\)-structure \(\cM\) is given by the subset of the elements of the structure in which the formula holds.
More precisely, if \(\varphi\) has free variables among the sequence of typed variables \(\vec{x} = (x_1:s_1,\ldots,x_n:s_n)\), the interpretation of \(\varphi\) is a subset \(S\) of \(M_{\vec{x}} = M_{s_1} \times \ldots \times M_{s_n}\) of values \(\vec{a} = (a_i)_{i\in\set{1,\ldots,n}}\), with \(a_i \in M_{s_i}\), satisfying \(\varphi\). The satisfaction of \(\varphi\) at the element \(\vec{a}\) is written \(\cM \models_{\vec{a}} \varphi\). Rephrasing the construction categorically in the category \(\Set\) of sets and functions, the subset \(S\) is obtained via the following pullback:
\[\begin{tikzcd}
	S & \terminal \\
	{M_{\vec{x}}} & {\set{0,1}}
	\arrow[from=1-1, to=1-2]
	\arrow["{ }", tail, from=1-1, to=2-1]
	\arrow["\lrcorner"{anchor=center, pos=0.125}, draw=none, from=1-1, to=2-2]
	\arrow["true", tail, from=1-2, to=2-2]
	\arrow["{\varphi(x)}", from=2-1, to=2-2]
\end{tikzcd}\]
where $\terminal = \set{*}$ is the terminal object in the category \(\Set\) and $\chi_S$ is the characteristic function associated with the inclusion \(S \subseteq M_{\vec{x}}\). A value \(\vec{a}\) then corresponds to a morphism \(\terminal \to M_{\vec{x}}\), meaning that the satisfaction of \(\varphi\) at \(\vec{a}\), i.e., \(\cM \models_{\vec{a}} \varphi\), is given by the commutativity of the following diagram:
\[
    \begin{tikzcd}
        & S \\
        \terminal & {M_{\vec{x}}.}
        \arrow["{ }", tail, from=1-2, to=2-2]
        \arrow["{ }", from=2-1, to=1-2]
        \arrow["{\vec{a}}", from=2-1, to=2-2]
    \end{tikzcd}
\]
Kripke--Joyal semantics generalizes this pointwise interpretation to the categorical framework by replacing the global elements of \(M_{\vec{x}}\), that is all morphisms \(\vec{a}\from \terminal \to M_{\vec{x}}\), with the generalized elements of \(M_{\vec{x}}\), that is all morphisms \(U \to M_{\vec{x}}\) (see~\cite[Sect. VI.6]{MM12} or~\cite[Chap. 5, Sect. 4]{johnstone_topos_1977}).
Kripke--Joyal semantics also admits rules, sometimes called semantic rules, explaining how connectives and quantifiers are to be interpreted using the notion of generalized elements. These rules correspond to \cref{thm:fundamental theorem of KJ semantics} in this paper, and are the key ingredient to proving our fundamental theorem (see Theorem~\ref{thm:Los:kjsem}) for Kripke--Joyal semantics.

The ultraproducts method has previously been studied abstractly in (fragments of) FOL~\cite{AN78,Dia08,Dia17}. To our knowledge, \Los's theorem has not been investigated in the setting of Kripke--Joyal semantics. A related categorical perspective is developed by Makkai~\cite{Mak87}, who construct small pretoposes from their categories of set-valued models equipped with an ultraproduct structure. 
Our approach is complementary:
we study ultraproducts of structures interpreted directly in the elementary topos and formulate \Los's theorem using its Kripke--Joyal semantics.

Our investigation for extending \Los's theorem to topoi highlights the need to address difficulties arising from the categorical setting. In particular, the interpretation of quantifiers, disjunction, and negation in Kripke--Joyal semantics involves local witnesses, while filtered products being filtered colimits have a global flavour. We prove that a locally presentable, two-valued topos with a projective terminal object provides sufficient conditions to control these interactions. From these hypotheses, we derive closure properties per connectives which entail a generalized \Los's theorem for increasingly expressive fragments of first-order logic. When the topos is assumed Boolean, we obtain a generalized \Los's theorem for the full first-order logic and recover the classical set-theoretic theorem (Theorem~\ref{thm:Los:set-theory}). We also derive compactness results for the associated logical fragments.

We also discuss the hypotheses that we used to prove the generalized forms of \Los's theorem. In particular, both \(\Set\times\Set\) and \(G\text{-}\Set\) satisfy \Los's theorem despite respectively failing two-valuedness and projectivity of \(\terminal\). This observation opens the question of a proper characterization of the topoi where \Los's theorem holds. We call this \emph{\Los's property} and, from the previous counterexamples, conjecture that \Los's property admits an external characterization in terms of conservative families of logical points compatible with the ultraproduct construction.

The paper is organized as follows. We introduce the notations used in the paper in~\cref{sec:preliminaries} and recall some theoretical backgrounds about topoi and filters. In~\cref{sec:categoricalFOL}, we present Kripke--Joyal semantics, i.e., a categorical semantics for FOL, while \cref{sec:los} is dedicated to our main positive result, \Los's theorem under local presentability, two-valuedness, and the projectivity of the terminal object.

\section{Preliminaries}
\label{sec:preliminaries}

We assume familiarity with the main notions from category theory, such as categories, functors, natural transformations, limits, colimits, and Cartesian closedness. We refer the interested reader to standard textbooks such as~\cite{BW90,McL71}. 

\subsection{Notations}
\label{subsec:preliminaries:notations}

Throughout the paper, we write \(\cC\) and $\cD$ for generic categories, $X$ and $Y$ for objects of categories, $\ob{\cC}$ for the collection of objects of a category \(\cC\), $f$, $g$, and $h$ for morphisms, \(\Hom_\cC(X,Y)\) for the hom-set from \(X\) to \(Y\) in \(\cC\), $F,G,H \from \cC \to \cD$ for functors from a category \(\cC\) into a category $\cD$, and $\alpha,\beta\from F \Rightarrow G$ for natural transformations.
Given a morphism $f\from X \to Y$, we write $\dom(f) = X$ for the domain of $f$, $\codom(f) = Y$ for its codomain, and $f\from X \mto Y$ if \(f\) is a monomorphism.
For an object \(X \in \ob{\cC}\), we write \(\id_X\) for the identity morphism on \(X\).
We write $\initial$ and $\terminal$ for the initial and terminal objects, \(X \times Y\) for the product of \(X\) and \(Y\) and \(X + Y\) for their coproduct, \(\prod_I X_i\) for the product of a family \((X_i)_{i \in I}\) and \(\coprod_I X_i\) for its coproduct.
Given two functors \(F\from \cC \to \cD\) and \(G\from \cD \to \cC\), \(F \dashv G\) means that \(F\) is left adjoint to \(G\).
Finally, when a category \(\cC\) is Cartesian closed, $X^Y$ denotes the exponential object of $X$ and $Y$.

\subsection{Subobjects and Heyting Algebras}
\label{subsec:preliminary:subobjects}

Elementary topoi generalize the category of sets and functions, allowing a more abstract solution for logical reasoning. While many approaches can be taken to present and define topoi~\cite{Johnstone02}, we view them as a structure of intuitionistic logic where the notion of truth value is deeply linked with that of subobjects. We first recall the notion of subobject before presenting topoi.

In a category \(\cC\), the set \(\Sub(X)\) of subobjects of an object \(X\) consists of equivalence classes of monomorphism into \(X\), where two monomorphisms \(f\from A \mto X\) and \(g\from B \mto X\) are equivalent if and only if \(A\) and \(B\) are isomorphic. We write \([f]\) for the equivalence class of \(f\). For instance, the subobjects in \(\Set\) of a set \(X\) are the subsets of \(X\) (up to isomorphism). Additionally, \(\Sub(X)\) admits a partial order \(\preceq_X\) such that \([f] \preceq_X [g]\) if \(f\) factors through \(g\), i.e., there is \(h \from A \mto B\) such that \(f = g \circ h\).

If \(\cC\) is a finitely complete category such that the poset \(\Sub(X)\) is viewed as a small category, the mapping \(X \mapsto \Sub(X)\) yields a contravariant functor \(\Sub\from \cC^{op} \to \Pos\), with \(\Pos\) being the category of posets. In addition to mapping objects \(X\) of \(\cC\) to \(\Sub(X)\), the functor maps morphisms \(f\from X \to Y\) to base change functors \(f^*\from \Sub(Y)\to \Sub(X)\). Given \(f\from X \to Y\), the base change functor \(f^*\) maps each \([Y' \mto Y]\) to \([X' \mto X]\), making the following diagram
\[\begin{tikzcd}
	{X'} & {Y'} \\
	X & Y
	\arrow[from=1-1, to=1-2]
	\arrow[tail, from=1-1, to=2-1]
	\arrow["\lrcorner"{anchor=center, pos=0.125}, draw=none, from=1-1, to=2-2]
	\arrow[tail, from=1-2, to=2-2]
	\arrow["f"', from=2-1, to=2-2]
\end{tikzcd}\]
a pullback.

For logical purposes, we want more structure than just posets for \(\Sub(X)\). Typically, we want \(\Sub(X)\) to be at least a Heyting algebra, i.e., a distributive bounded lattice that
admits an implication $\to$ such that for any \(A\) in \(\Sub(X)\), the following adjunction holds \((\_ \wedge A) \dashv (A \to \_)\). The largest and smallest objects of \((\Sub(X),\preceq_X)\) are respectively \([\id_X]\) and \([\initial \mto X]\).

For simplicity, we may write \(Y \mto X\) instead of \([Y \mto X]\) for a subobject, identifying the equivalence class with a representative element, or even \(Y\) when there is no ambiguity on the codomain \(X\).

\subsection{Elementary Topoi}
\label{subsec:preliminaries:topos}

Elemantary topoi are categories that allow equipping the poset \(\Sub(X)\) with such a structure of Heyting algebra.
Topoi are finitely complete Cartesian closed category with a subobject classifier~\cite[Chap. A2]{Johnstone02}. A subobject classifier is a monomorphism \(true\from \terminal \mto \Omega\) out of the terminal object \(\terminal\), such that for every monomorphism \(m\from Y \mto X\), there exists a unique morphism \(\chi_m\from X \to \Omega\) such that the following diagram is a pullback:
\[\begin{tikzcd}
	Y & \terminal \\
	X & \Omega
	\arrow["{!}", from=1-1, to=1-2]
	\arrow["m"', tail, from=1-1, to=2-1]
	\arrow["\lrcorner"{anchor=center, pos=0.125}, draw=none, from=1-1, to=2-2]
	\arrow["true", tail, from=1-2, to=2-2]
	\arrow["{\chi_m}"', from=2-1, to=2-2]
\end{tikzcd}\]
The morphism \(\chi_m\) is called the {\bf characteristic} or {\bf classifying morphism} of \(m\). Hence, $\Omega$ represents the contravariant functor $\Sub$, i.e., for every $X \in Ob(\cC)$, $\Sub(X) \simeq \Hom_\cC(X,\Omega)$, and the universal object is $true\from \terminal \to \Omega$.

A topos also fulfills the following properties~\cite{BW85,Johnstone02}:

\begin{itemize}
	\item It is finitely cocomplete.
    \item It has an initial object $\initial$ and a terminal object $\terminal$, which are the colimit and limit of the empty diagram (since it is finitely complete and cocomplete).
    \item Epimorphisms and monomorphisms form a factorization system, i.e., every morphism $f$ can be uniquely factorized as $m_f \circ e_f$ where $e_f$ is an epimorphism and $m_f$ is a monomorphism. The codomain of $e_f$ is called the {\bf image of} $f$ and is written $\im(f)$. Then $(A \stackrel{f}{\rightarrow} B) = (A \stackrel{e_f}{\rightarrow} \im(f) \stackrel{m_f}{\mto} B)$.
	\item Every object \(X\) in \(\ob{\cC}\) has a {\bf power object} \(PX\) defined as the exponential \(\Omega^X\). 
    Power objects generalize the notion of powerset from the category \(\Set\).
    As a topos is Cartesian closed, a power object satisfies the following adjunction: 
	\[
        \Hom_\cC(X \times Y,\Omega) \simeq \Hom_\cC(X,PY).
    \]
\end{itemize}

Topoi encompass \(\Set\), categories of presheaves, and more generally Grothendieck topoi, i.e., categories equivalent to the category of sheaves over a site~\cite{Car17}.

In a topos, the poset of subobjects \(\Sub(X)\) is a Heyting algebra~\cite{Johnstone02}. Additionally, each base change functor \(f^*\) admits both left and right adjoints $\exists_f$ and $\forall_f$, i.e., \(\exists_f \dashv f^* \dashv \forall_f\).

In \cref{thm:Los:kjsem}, we will consider two special cases of topoi: boolean and two-valued topos. A topos is {\bf Boolean} if \(\Sub(X)\) is a Boolean algebra for every object \(X\), meaning that every subobject has a complement. A topos is {\bf two-valued} if it admits exactly two truth values. More precisely, its subobject classifier \(\Omega\) admits two global elements \(\mathrm{true}\from \terminal \to \Omega\) and \(\mathrm{false}\from \terminal \to \Omega\), or equivalently if the terminal object \(\terminal\) has exactly two subobjects, \(\initial \mto \terminal\) and \(\id_\terminal\).

\subsection{Filters, Filtered Products, and Filtered Colimits}
\label{subsec:preliminaries:filters}

\Los's theorem relies on filters and ultrafilters, which give rise to categorical diagrams whose colimits define filtered products. The construction admits a natural generalization to filtered colimits, which in turn, lead to the concept of presentability. This concept provides the appropriate framework for the categorical generalization of \Los's theorem.

\subsubsection{Set-theoretic Filters and Ultrafilters}
\label{subsec:preliminaries:filters:set}

Given a nonempty set $I$, a {\bf filter} $F$ {\bf over} $I$ is a subset of $\powerset(I)$ such that:\footnote{$\powerset(I)$ denotes the powerset of $I$.}
\begin{itemize}
	\item $I \in F$;
	\item if $A,B \in F$, then $A \cap B \in F$, and
	\item if $A \in F$ and $A \subseteq B$, then $B \in F$.
\end{itemize}

For instance, \(\set{I}\) and \(\powerset(I)\) are filters on \(I\). The filter generated by some $J \subseteq I$ is $F_J = \set{A \in \powerset(I) \mid J \subseteq A}$. It is called a {\bf principal filter}. If $I$ is finite, all filters on $I$ are principal.

A filter $F$ is {\bf proper} when $F$ is not $\powerset(I)$ and we later assume all filters to be proper. A filter is an {\bf ultrafilter} if it is maximal for inclusion. In particular, if \(U\) is an ultrafilter, then every $A \in \powerset(I)$ is in $U$ if and only if $I \setminus A$ is not in $U$. By Zorn's lemma, any filter is contained in an ultrafilter.

\subsubsection{Filtered Products in Categories}
\label{subsec:preliminaries:filters:products}

In \(\Set\), filtered products correspond to directed colimits of products of sets, which have then been extended to arbitrary categories~\cite{Okh66}. The idea is to consider a filter \(F\) over a set \(I\) as a category (ordered by reverse inclusion) such that a filtered product can be defined as the colimit of the directed diagram of projections between the products over the elements of the filter~\cite{AN78,Dia08,Dia17}.

\begin{Definition}[Filtered product]
\label{def:filteredproduct}
Let \(F\) be a filter over a set of indices $I$, and let $X = (X_i)_{i \in I}$ be a $I$-indexed family of objects in \(\cC\). Then, \(F\) and \(X\) induce a functor $A_F\from F^{op} \to \cC$, mapping each subset inclusion $J \subseteq J'$ of $F$ to the canonical projection $p_{J',J}\from \prod_{J'} X_j \to \prod_J X_j$.

The {\bf filtered product} of $X$ modulo \(F\) is the colimit $\mu\from A_F \Rightarrow \prod_F X$ of the functor $A_F$.
\(\cC\) {\bf have filtered products} if any filter \(F\) and any \(I\)-indexed family of objects \(X = (X_i)_{i \in I}\) in \(\cC\) yield a filtered product of \(X\) modulo \(F\).
\end{Definition}

Filtered products are unique up to isomorphisms since they are colimits, and we can talk about \emph{the} filtered product of \(X\) modulo \(F\).

For instance, \(\Set\) have filtered products, and, therefore, any presheaf topos \(\widehat{\cB}\) for some small category \(\cB\) also have them. Indeed, presheaf limits and colimits are computed componentwise, meaning that filtered products can be lifted from sets to presheaves.

\subsubsection{Locally Presentable Category}
\label{subsec:preliminaries:filters:lfp}

We now introduce filtered colimits and presentable objects, following the presentation given in the reference book~\cite{adamek_locally_1994}. These notions generalize the construction of filtered products introduced above. We consider a regular cardinal \(\lambda\), that is an infinite cardinal which is not a sum of a smaller number of smaller cardinals.

The notion of filter viewed as a category (\(F^{op}\) in Definition~\ref{def:filteredproduct}) can be generalized to obtain the notion of filtered category.

\begin{Definition}[Filtered category]
    A category \(\cJ\) is {\bf filtered} if every finite diagram in \(\cJ\) admits a cocone. 
\end{Definition}

Equivalently, \(\cJ\) is filtered if 
\begin{itemize}
    \item for any two objects \(i,j\), there exists \(k\) and morphisms \(i \to k\) and \(j \to k\), and
    \item for any parallel morphisms \(f,g\from i \rightrightarrows j\), there exists \(k\) and a morphism \(h\from j \to k\) such that \(h \circ f = h \circ g\).
\end{itemize}
The definition naturally extends to a regular cardinal \(\lambda\).

\begin{Definition}[\(\lambda\)-filtered category]
    A category \(\cJ\) is {\bf \(\lambda\)-filtered} if every diagram of size strictly less than \(\lambda\) admits a cocone.
\end{Definition}

In particular, filtered categories are exactly \(\aleph_0\)-filtered categories. Filtered and \(\lambda\)-filtered categories serve as indices to define filtered and \(\lambda\)-filtered colimits.

\begin{Definition}[\(\lambda\)-filtered colimit]
    A {\bf \(\lambda\)-filtered colimit} is a colimit of a functor \(D\from \cJ \to \cC\) where the shape (or index category) \(\cJ\) of \(D\) is \(\lambda\)-filtered.
\end{Definition}

In the special case where \(\lambda\) is \(\aleph_0\), the colimit is simply said to be a filtered colimit. Filtered colimits generalize filtered products: the diagram \(A_F\from F^{op} \to \cC\) from Definition~\ref{def:filteredproduct} is indexed by a filtered category. 
Filtered and \(\lambda\)-filtered colimits play a central role in defining presentability.

\begin{Definition}[Presentable object]
	\label{def:fp object}
    Let \(\cC\) be a category. An object \(X\) in \(\cC\) is {\bf \(\lambda\)-presentable} if the hom-functor \(\Hom_\cC(X, \_)\from \cC \to \Set\) preserves \(\lambda\)-filtered colimits.
    \(X\) is {\bf presentable} if it is \(\lambda\)-presentable for some regular cardinal \(\lambda\) and {\bf finitely presentable} if it is \(\aleph_0\)-presentable.
\end{Definition}

Unfolding this definition, for any filtered diagram \(D\from \cJ \to \cC\) with colimit cocone \(\nu\from D \Rightarrow V\), every morphism \(\mu\from X \to V\) in \(\cC\) factors essentially uniquely through some stage of the diagram. More precisely:
\begin{itemize}
    \item there exist \(i\) in \(\cJ\) and \(\mu_i\from X \to D(i)\) such that \(\mu = \nu_i \circ \mu_i\);
    \item any two such factorizations become equal at some later stage of the diagram, that is, for any two morphisms \(\mu_i\) and \(\mu_j\) as above, there exist \(k\) in \(\cJ\), with morphisms morphisms \(i \to k\) and \(j \to k\) such that \(D(i \to k) \circ \mu_i = D(j \to k) \circ \mu_j\).
\end{itemize}

\begin{Definition}[Locally presentable category]
	\label{def:lp category}
    A category \(\cC\) is {\bf locally \(\lambda\)-presentable} if
    \begin{itemize}
        \item it is cocomplete, and
        \item there exists a set \(A\) of \(\lambda\)-presentable objects, called {\bf generators}, such that every object in \(\cC\) is a \(\lambda\)-filtered colimit of objects in \(A\).
    \end{itemize}
    The category is {\bf locally presentable} if it is locally \(\lambda\)-presentable for some regular cardinal \(\lambda\) and {\bf locally finitely presentable} if locally \(\aleph_0\)-presentable.
\end{Definition}

Locally presentable categories are in fact complete and cocomplete, meaning that they admit all small limits in addition to the colimits assumed in the definition. Important examples include presheaf topoi and sheaf topoi~\cite[Chap.~D2, Sect.~3]{Johnstone02}. Additional examples can be consulted in~\cite{adamek_locally_1994}.

We conclude with
two lemmata, related to the behavior of some properties under filtered colimits. First,
a lemma stating that, in locally finitely presentable categories, filtered colimts commute with finite limits~\cite[Theorem 1 (p. 215), Sect. 2, Chap. IX]{McL71}, which generalizes to \(\lambda\)-filtered colimits and \(\lambda\)-small limits in locally presentable categories~\cite[Prop. 1.59]{adamek_locally_1994}:

\begin{Lemma}
\label{lma:filtered-colimits-commutes-small-limits}
In a locally \(\lambda\)-presentable category, \(\lambda\)-filtered colimits commute with \(\lambda\)-small limits.
\end{Lemma}

Second, every cocone morphism of a filtered colimit whose colimit maps are epimorphisms is itself an epimorphism.

\begin{Lemma}
\label{lma:colimit-cocone-epi}
Let \(D \from \cJ \to \cC\) be a filtered diagram in a category \(\cC\), and let \((\iota_j \from D(j) \to C)_{j \in \cJ}\) be a colimit cocone. If every morphism of the diagram \(D(u)\) is an epimorphism, then each cocone morphism \(\iota_j\) is an epimorphism.
\end{Lemma}

\begin{proof}
Let \(j \in \cJ\), and let \(f,g \from C \to X\) be morphisms such that \(f \circ \iota_j = g \circ \iota_j\).

Fix \(k \in \cJ\). Since \(\cJ\) is filtered, there exist an object \(\ell\) and two morphisms \(u \from j \to \ell\) and \(v \from k \to \ell\). Now, \(\iota_\ell \circ D(u) = \iota_j\) and \(\iota_\ell \circ D(v) = \iota_k\), such that 
\[
f \circ \iota_\ell \circ D(u) = g \circ \iota_\ell \circ D(u).
\]
Because \(D(u)\) is an epimorphism, it can be cancelled. Composing with \(D(v)\), it follows that \(f \circ \iota_k = g \circ \iota_k\).

Since this holds for every \(k \in \cJ\), the universal property of the colimit implies \(f=g\). Hence \(\iota_j\) is an epimorphism.
\end{proof}

\section{Categorical First-Order Logic (FOL)}
\label{sec:categoricalFOL}

Following the seminal work of Alfred Tarski~\cite{Tar44,Tar56}, it is well-established that FOL can be interpreted with set-theoretic models where sorts are interpreted as sets, function symbols as functions between sets, and relation symbols as subsets of Cartesian products. Categorical logic builds on this construction, interpreting sorts as objects, terms as morphisms, and formulae as subobjects. In this section, we consider an elementary topos \(\cC\).

\subsection{The Category of \mathinhead{\Sigma}{Sigma}-Structures}
\label{subsec:categoricalFOL:FOLstructures}

In the sequel, we consider a FOL-signature \(\Sigma = (S, F, R)\), where \(S\) is the set of sorts, \(F\) the set of function symbols, and \(R\) the set of relation symbols. We first recall the notion of \(\Sigma\)-structures in a topos~\cite[Chap. D1]{Johnstone02}.

\begin{Definition}[$\Sigma$-structure]
A {\bf $\Sigma$-structure} \(\cM\) in \(\cC\) is defined by:
\begin{itemize}
	\item an object $M_s \in \ob{\cC}$ for every sort \(s\) in \(S\), 
	\item a morphism $f^\cM\from M_{s_1} \times \ldots \times M_{s_n} \to M_s$ in \(\cC\) for every function symbol \(f\) in \(F\) with profile \(s_1 \times \ldots \times s_n \to s\), and \(f^\cM\from \terminal \to M_s\) if \(f\) is a constant symbol (i.e., \(n = 0\)),
	\item a subobject $r^\cM \in \Sub(M_{s_1} \times \ldots \times M_{s_n})$ for every relation symbol \(r\) in \(R\) with profile \(s_1 \times \ldots \times s_n\). 
\end{itemize}

A {\bf $\Sigma$-structure morphism} $h\from \cM \to \cN$ in \(\cC\) is a family of morphisms $(h_s\from M_s \to N_s)_{s \in S}$ such that:
\begin{itemize}
	\item the diagram
	\[\begin{tikzcd}
        {M_{s_1} \times \ldots \times M_{s_n}} & {M_s} \\
        {N_{s_1} \times \ldots \times N_{s_n}} & {N_s}
        \arrow["{f^\cM}", from=1-1, to=1-2]
        \arrow["{h_s}", from=1-2, to=2-2]
        \arrow["{h_{s_1} \times \ldots \times h_{s_n}}"', from=1-1, to=2-1]
        \arrow["{f^\cN}"', from=2-1, to=2-2]
    \end{tikzcd}\]
	commutes, for every function symbol \(f\from s_1 \times \ldots \times s_n \to s\) in \(F\).

	\item there is a morphism \(O \to O'\) such that the diagram
	\[\begin{tikzcd}
        O & {M_{s_1} \times \ldots \times M_{s_n}} \\
        {O'} & {N_{s_1} \times \ldots \times N_{s_n}}
        \arrow["{h_{s_1} \times \ldots \times h_{s_n}}", from=1-2, to=2-2]
        \arrow["{r^\cM}", tail, from=1-1, to=1-2]
        \arrow["{r^\cN}"', tail, from=2-1, to=2-2]
        \arrow[from=1-1, to=2-1]
    \end{tikzcd}\]
    commutes, for every relation symbol \(r\from s_1 \times \ldots \times s_n\) in \(R\). 
\end{itemize}

\(\Sigma\)-structures and \(\Sigma\)-structure morphisms in \(\cC\) form a category written \(\Structures{\cC}\).
\end{Definition}

\begin{Proposition}
\label{prop:has small products}
If \(\cC\) has small products, then so does $\Structures{\cC}$.
\end{Proposition}

\begin{proof}
Let \(I\) be a set and \((\cM_i)_{i \in I}\) be an \(I\)-indexed family of models. Consider the model $\prod_I \cM_i$ defined by:
\begin{itemize}
    \item for every $s \in S$, \((\prod_I M_i)_s = \prod_I (M_i)_s\), with \(\prod_I (M_i)_s\) being the product in \(\cC\),
    \item for every $f\from s_1 \times \ldots \times s_n \rightarrow s$ in $F$, $f^{\prod_I \cM_i}$ is the unique morphism such that the following diagram
    \[\begin{tikzcd}[column sep=4em]
    	{(\prod_I M_i)_{s_1} \times \ldots \times (\prod_I M_i)_{s_n}} & {(\prod_I M_i)_s} \\
    	{(M_i)_{s_1} \times \ldots \times (M_i)_{s_n}} & {(M_i)_s}
    	\arrow["{{f^{\prod_I \cM_i}}}", from=1-1, to=1-2]
    	\arrow["{(p_{I,i})_{s_1} \times \ldots \times (p_{I,i})_{s_n}}"', from=1-1, to=2-1]
    	\arrow["{(p_{I,i})_s}", from=1-2, to=2-2]
    	\arrow["{{f^\cM_i}}"', from=2-1, to=2-2]
    \end{tikzcd}\]
    commutes for all \(i \in I\), which is well-defined by the universal property of small products in \(\cC\),
    \item for every $r\from s_1 \times \ldots \times s_n$ in $R$, $r^{\prod_I \cM_i}$ is the subobject $\prod_I O_i \mto (\prod_I M_i)_{s_1} \times \ldots \times (\prod_I M_i)_{s_n}$ where $r^{\cM_i}\from O_i \mto (M_i)_{s_1} \times \ldots \times (M_i)_{s_n}$.
\end{itemize}
Since each \((\prod_I M_i)_s\) for \(s \in S\) is obtained as a small product in \(\cC\), it follows that $\prod_I \cM_i$ is the small product of $(\cM_i)_{i \in I}$.
\end{proof}

If \(x\) is a $\Sigma$-variable of sort \(s\), we write \(x\from s\). We also write \(\vec{x}\) for a sequence of variables and \(\vec{x_1}\vec{x_2}\), resp. \(\vec{x_1}y\), when concatenating such sequences, resp. appending a new variable. These two notations naturally extend to terms. We say that a sequence of variables $\vec{x} = (x_1:s_1,\ldots,x_n:s_n)$ is a {\bf suitable context} for a $\Sigma$-term, resp. a $\Sigma$-formula, if all free variables of this term, resp. formula, belong to $\set{x_1,\ldots,x_n}$. We write $\vec{x}.t$, resp. $\vec{x}.\varphi$, to denote that $\vec{x}$ is a suitable context for $t$, resp. $\varphi$. Then $\vec{x}.t$, resp. $\vec{x}.\varphi$, is called a {\bf term-in-context}, resp. a {\bf formula-in-context}. Additionally, we write \(M_{\vec{x}}\) instead of \(M_{s_1} \times \ldots \times M_{s_n}\) for a sequence of variables \(\vec{x} = (x_1:s_1,\ldots,x_n:s_n)\).

Given a $\Sigma$-structure \(\cM\), the {\bf interpretation} of a term-in-context $\vec{x}.t$ in \(\cM\), with \(t:s\), is a morphism \(\sem[\cM]{\vec{x}.t} \from M_{\vec{x}} \to M_s\), defined inductively as follows:
\begin{itemize}
    \item if \(t\) is a variable, then it corresponds to an \(x_j\) in \(\vec{x}\), and \(\sem[\cM]{\vec{x}.x_j}\) is the projection on the sort associated with \(x_j\);
    \item if \(t\) is \(f(t_1, \ldots t_m)\) for some function symbol \(f\) and \(t_1\from s'_1, \ldots, t_m\from s'_m\), then \(\sem[\cM]{\vec{x}.t}\) is the composite
    \[
        M_{\vec{x}} \xrightarrow{(\sem[\cM]{\vec{x}.t_1},\ldots,\sem[\cM]{\vec{x}.t_m})} M_{s'_1} \times \ldots \times M_{s'_m} \xrightarrow{f^\cM} M_{s}.
    \]
\end{itemize}

The interpretation of terms-in-context extends to the interpretation of formulae-in-context $\vec{x}.\varphi$ in \(\cM\), written $\sem[\cM]{\vec{x}.\varphi}$ and defined inductively as follows:

\begin{itemize}
\item if \(\varphi\) is $\bot$, then $\sem[\cM]{\vec{x}.\bot}$ is the least element of $\Sub(M_{\vec{x}})$, i.e, \(\initial \to M_{\vec{x}}\) ;
\item if \(\varphi\) is $\top$, then $\sem[\cM]{\vec{x}.\top}$ is the greatest element of $\Sub(M_{\vec{x}})$, i.e., \(\id_{M_{\vec{x}}}\) ;
\item if \(\varphi\) is an equation $t = t'$, then $\sem[\cM]{\vec{x}.(t = t')}$ is the equalizer of
\[\begin{tikzcd}[column sep=5em]
    {M_{\vec{x}}} & {M_s}
    \arrow["{\sem[\cM]{\vec{x}.t'}}"', shift right, from=1-1, to=1-2]
    \arrow["{\sem[\cM]{\vec{x}.t}}", shift left, from=1-1, to=1-2]
\end{tikzcd}\]
where \(s\) is the common sort of \(t\) and \(t'\);
\item if \(\varphi\) is a relation \(r(\vec{t})\) with \(r\from s_1 \times \ldots \times s_n\), then \(\sem[\cM]{\vec{x}.r(\vec{t})}\) is the subobject \(O'\mto M_{\vec{x}}\) obtained from the following pullback:
\[\begin{tikzcd}
	{O'} & O \\
	{M_{\vec{x}}} & {M_{s_1} \times \ldots \times M_{s_n}}
	\arrow[from=1-1, to=1-2]
	\arrow["{\sem[\cM]{\vec{x}.r(\vec{t})}}"', tail, from=1-1, to=2-1]
	\arrow["\lrcorner"{anchor=center, pos=0.125}, draw=none, from=1-1, to=2-2]
	\arrow["{r^\cM}", tail, from=1-2, to=2-2]
	\arrow["{\sem[\cM]{\vec{x}.\vec{t}}}"', from=2-1, to=2-2]
\end{tikzcd}\]
\item if \(\varphi\) is a formula of the form $\psi \wedge \chi$, $\psi \lor \chi$, or $\psi \Rightarrow \chi$, then \(\sem[\cM]{\vec{x}.\varphi}\) is interpreted via the operators $\wedge$, $\lor$, and $\to$ of the Heyting algebra $\Sub(M_{\vec{x}})$.
\item if \(\varphi\) is a formula of the form \(\exists y.\psi\), with $y:s$, then $\sem[\cM]{\vec{x}.(\exists y.\psi)} = \exists_\pi \sem[\cM]{\vec{x}y.\psi}$ with $\pi$ being the projection $\from M_{\vec{x}} \times M_s \to M_{\vec{x}}$ and $\exists_\pi \dashv \pi^*$.
\item if \(\varphi\) is a formula of the form \(\forall y.\psi\) then $\sem[\cM]{\vec{x}.(\forall y.\psi)} = \forall_\pi \sem[\cM]{\vec{x}y.\psi}$, with $\pi$ being the same projection and $\pi^* \dashv \forall_\pi$.
\end{itemize}

Note that, by construction, the interpretation \(\sem[\cM]{\vec{x}.\varphi}\) of the formula-in-context \(\vec{x}.\varphi\) in \(\cM\) is a subobject of \(M_{\vec{x}}\). We write \(\set{\vec{x} \mid \varphi(\vec{x})}_\cM\) for its domain, i.e., \(\sem[\cM]{\vec{x}.\varphi}\from \set{\vec{x} \mid \varphi(\vec{x})}_\cM \mto M_{\vec{x}}\).

A FOL formula is a {\bf sentence} if \(\emptycontext\) (the empty sequence of variables) is a suitable context for it, meaning that its interpretation in a \(\Sigma\)-structure \(\cM\) is a morphism with codomain \(M_{{\emptycontext}}\), that is the terminal object \(\terminal\) of \(\cC\). 

\subsection{Kripke--Joyal Semantics}
\label{subsec:categoricalFOL:KripkeJoyal}

In most categories, the elements of an object do not correspond to a well-defined notion. Therefore, the notion of elements of a \(\Sigma\)-structure needs to be replaced to redefine the satisfaction of FOL formulae. A solution is to consider the generalized elements of \(\cC\), thus regarding \(\cC\) as its image under the Yoneda embedding. Indeed, by Yoneda's lemma, an object \(X\) is uniquely determined, up to isomorphisms, by the functor \(\Hom_\cC(\_, X)\), i.e., the morphisms \(Y \to X\) in \(\cC\), also called {\bf generalized elements}~\cite{Car17}. In topoi, the fundamental theorem~\cite{mclarty_elementary_1992}, or slice theorem, ensures that the generalized elements of \(X\) in the topos \(\cC\) correspond to ordinary points in the slice topos \(\cC/X\).

In this subsection, we consider an object \(\cM\) of \(\Structures{\cC}\), i.e., a \(\Sigma\)-structure in \(\cC\), a \(\Sigma\)-formula \(\varphi\) with a suitable context \(\vec{x} = (x_1:s_1,\ldots,x_n:s_n)\), and a generalized element \(\alpha\from U \to M_{\vec{x}}\).

Kripke--Joyal semantics is thoroughly explained in~\cite[Sect. VI.6]{MM12}, with different notations and using the forcing relation \(U \Vdash \varphi(\alpha)\) between the generalized element and the formula, instead of a relation between the \(\Sigma\)-structure and the formula. More compact explanations might be found in~\cite[Chap. 5, Sect. 4]{johnstone_topos_1977}, again using different notations.

\begin{Definition}[Kripke--Joyal semantics]
\label{def:KJsem}
The satisfaction of \(\vec{x}.\varphi\) in \(\cM\) by the generalized element \(\alpha\), written \(\cM \models_\alpha \vec{x}.\varphi\), is defined as:
\[
    \cM \models_\alpha \vec{x}.\varphi \; \mbox{iff} \; \alpha \; \mbox{factors through} \; \sem[\cM]{\vec{x}.\varphi}
\]
i.e., the following diagram commutes:

\[\begin{tikzcd}
	& {\set{\vec{x} \mid \varphi(\vec{x})}_\cM} & \terminal \\
	U & {M_{\vec{x}}} & {\Omega.}
	\arrow[from=1-2, to=1-3]
	\arrow["{\sem[\cM]{\vec{x}.\varphi}}", tail, from=1-2, to=2-2]
	\arrow["true", tail, from=1-3, to=2-3]
	\arrow["{ }", from=2-1, to=1-2]
	\arrow["\alpha", from=2-1, to=2-2]
	\arrow["{\varphi(x)}", from=2-2, to=2-3]
\end{tikzcd}\]

Equivalently, this means that $\im(\alpha) \preceq \set{\vec{x} \mid \varphi(\vec{x})}_\cM$.
\end{Definition}

As discussed in~\cite[Chap. VI, Sect. 6]{MM12}, Kripke--Joyal semantics satisfies the following two properties, called {\bf monotonicity} and {\bf local character}.

\begin{Proposition}[Monotonicity~{\protect\cite[Chap.VI, Sect.6]{MM12}}]
\label{prop:monotonicity}
If \(\cM \models_\alpha \vec{x}.\varphi\), then, for any morphism \(f \from V \to U\) in \(\cC\), \(\cM \models_{\alpha \circ f} \vec{x}.\varphi\).
\end{Proposition}

\begin{Proposition}[Local character~{\protect\cite[Chap.VI, Sect.6]{MM12}}]
\label{prop:local character}
If \(f \from V \to U\) is an epimorphism and \(\cM \models_{\alpha \circ f} \vec{x}.\varphi\), then \(\cM \models_\alpha \vec{x}.\varphi\).
\end{Proposition}

These two properties are the key to proving the following theorem, allowing for an inductive description of Kripke--Joyal semantics. This inductive description will later turn useful as the proof of \Los's theorem is conducted by induction. This theorem is also known as the semantic rules of Kripke--Joyal semantics.

\begin{Theorem}[Theorem VI.6.1 in~\cite{MM12}]
\label{thm:fundamental theorem of KJ semantics}
Let $\alpha\from U \to M_{\vec{x}}$ be a generalized element, $\varphi$ and $\psi$ be a \(\Sigma\)-formula, then
\begin{itemize}
	\item $\cM \models_\alpha \vec{x}.\varphi \wedge \psi$ iff $\cM \models_\alpha \vec{x}.\varphi$ and $\cM \models_\alpha \vec{x}.\varphi$;
	\item $\cM \models_\alpha \vec{x}.\varphi \lor \psi$ iff there are morphisms $p\from V \to U$ and $q\from W \to U$ such that $p+q\from V + W \to U$ is an epimorphism, and both $\cM \models_{\alpha \circ p} \vec{x}.\varphi$ and $\cM \models_{\alpha \circ q} \vec{x}.\psi$;

	\item $\cM \models_\alpha \vec{x}.\varphi \Rightarrow \psi$ iff for any morphism $p\from V \to U$ such that $\cM \models_{\alpha \circ p} \vec{x}.\varphi$, then $\cM \models_{\alpha \circ p} \vec{x}.\psi$;
	\item $\cM \models_\alpha \vec{x}.\neg \varphi$ iff for any morphism $p\from V \to U$ such that $\cM \models_{\alpha \circ p} \vec{x}.\varphi$, then $V \simeq \emptyset$;
\end{itemize}
For the quantifiers, we consider an additional variable $y:s$. Then
\begin{itemize}
	\item $\cM \models_\alpha \vec{x}.(\exists y.\varphi)$ iff there exists an epimorphism $p\from V \to U$ and a generalized element $\beta\from V \to M_s$ such that $\cM \models_{(\alpha \circ p,\beta)} \vec{x}y.\varphi$;
	\item $\cM \models_\alpha \vec{x}.(\forall y.\varphi)$ iff for every morphism $p\from V \to U$ and every generalized element $\beta\from V \to M_s$, it holds that $\cM \models_{(\alpha \circ p,\beta)} \vec{x}y.\varphi$;
\end{itemize}
\end{Theorem}

Kripke--Joyal semantics provide a notion of satisfaction relative to a generalized element, which can be aggregated into a global notion.

\begin{Definition}[Model]
    \label{def:model}
    \(\cM\) is a {\bf model for} \(\vec{x}.\varphi\), written \(\cM \models \vec{x}.\varphi\), if for all generalized elements \(\alpha\from U \to M_{\vec{x}}\), \(\cM \models_\alpha \vec{x}.\varphi\).
\end{Definition}

\begin{Proposition}
    \label{prop:satisfiedbyid}
    \(\cM \models \vec{x}.\varphi\) if and only if \(\cM \models_{\id_{M_{\vec{x}}}} \vec{x}.\varphi\).
\end{Proposition}
\begin{proof}
    The direct implication is obvious, and the converse follows from monotonicity (Proposition~\ref{prop:monotonicity}). 
\end{proof}

\subsection{Kripke--Joyal Semantics in Locally Presentable Topoi}
\label{subsec:categoricalFOL:presentability}

Kripke--Joyal semantics applies to arbitrary elementary topoi. However, for the developments that follow, in particular the categorical extension of \Los's theorem, the appropriate settings are locally presentable topoi, where Kripke--Joyal satisfaction can be tested on presentable generalized elements only.

\begin{Proposition}
\label{prop:validation restricted}
    If \(\cC\) is a locally presentable elementary topos, then \(\cM \models \vec{x}.\varphi\) if and only if for all generalized elements \(\alpha\from U \to M_{\vec{x}}\) such that \(U\) is presentable \(\cM \models_\alpha \vec{x}.\varphi\).
\end{Proposition}

\begin{proof}
    The direct implication is immediate. For the converse, let \(\alpha\from U \to M_{\vec{x}}\) be a generalized element. Since \(\cC\) is locally presentable, \(U\) is a filtered colimit of presentable objects \((A_i)_{i \in I}\) (see Definition~\ref{def:lp category}) with coprojections \(\nu_i \from A_i \to U\). 
    For each \(i\in I\), \(\alpha \circ \nu_i \from A_i \to M_{\vec{x}}\) is a generalized element with presentable domain. By hypothesis, \(\cM \models_{\alpha \circ \nu_i} \vec{x}.\varphi\), i.e., \(\alpha \circ \nu_i\) factors through \(\sem[\cM]{\vec{x}.\varphi}\) (see Definition~\ref{def:KJsem}).
    By the universal property of colimits, there is a unique morphism \(U \to {\set{\vec{x} \mid \varphi(\vec{x})}_\cM}\) making the relevant diagram commute. Thus, \(\alpha\) factors through \(\sem[\cM]{\vec{x}.\varphi}\), i.e., \(\cM \models_\alpha \vec{x}.\varphi\). From Definition~\ref{def:model}, we conclude that \(\cM \models \vec{x}.\varphi\).
\end{proof}

When the topos is locally presentable, it also follows that the witness for existential statements can be chosen over presentable objects.

\begin{Proposition}
\label{prop:existential and representable object}
Let \(\cC\) be a locally presentable elementary topos and $\alpha\from U \to M_{\vec{x}}$ a generalized element with presentable domain $U$. Then, $\cM \models_\alpha \vec{x}.(\exists y.\varphi)$ iff there exists an epimorphism $p\from V \to U$ and a generalized element $\beta\from V \to M_s$, with \(V\) presentable, such that $\cM \models_{(\alpha \circ p,\beta)} \vec{x}y.\varphi$.
\end{Proposition}

The proof rests on the following lemma.

\begin{Lemma}
    \label{lemma:existential and representable object}
    Let \(\cC\) be a \(\lambda\)-locally presentable topos. Let $f\from V \to U$ be an epimorphism such that $U$ is a \(\lambda\)-presentable object. Then, there exists a \(\lambda\)-presentable object $V'$ and a morphism $g\from V' \to V$ such that $f \circ g\from V' \to U$ is an epimorphism.
\end{Lemma}

\begin{proof}[Proof of the lemma]
    Let \(\cC\) be a \(\lambda\)-locally presentable topos. Let \(f\from V \to U\) be an epimorphism such that \(U\) is a \(\lambda\)-presentable object.

    Since \(\cC\) is locally presentable, we can write \(V = \colim_{i\in I} V_i\) with \(I\) a \(\lambda\)-filtered category and \(V_i\) \(\lambda\)-presentable objects of \(\cC\). Write \(\nu_i \from V_i \to V\) for the structure morphisms and define, by composition, morphisms \(h_i = f \circ \nu_i \from V_i \to U\). By the universal property of colimits, the cocone \((h_i)_{i\in I}\) induces a unique morphism \(\colim_{i\in I} V_i \to U\), which identifies with \(f\).
    Besides, \(f\) is an epimorphism and \(\im(f) \simeq U\).
    In a locally \(\lambda\)-presentable category, \(\lambda\)-filtered colimits commute with \(\lambda\)-small limits (Lemma~\ref{lma:filtered-colimits-commutes-small-limits}). Additionally, as filtered colimits are exact in a locally presentable topos, they preserve regular epimorphisms, they also preserve images. It follows that
    \[
        U \simeq \im(f) = \im(\colim_{i \in I} h_i) \simeq \colim_{i \in I} \im(h_i).
    \]

    Now, since \(U\) is \(\lambda\)-presentable, the identity \(\id_U\) factors through one of the coprojections \(\im(h_i) \mto U\), say \(u_j\from \im(h_j) \mto U\). Thus, there exists a morphism \(x\from U \to \im(h_j)\) such that \(\id_U = u_j \circ x\). Thus \(u_j\) is an isomorphism, and, in particular, \(h_j = f \circ \nu_j \from V_j \to U\) is an epimorphism.
\end{proof}

\begin{proof}[Proof of Proposition~\ref{prop:existential and representable object}]
    The converse implication is immediate. The directe one follows from Theorem~\ref{thm:fundamental theorem of KJ semantics} and Lemma~\ref{lemma:existential and representable object} that allows replacing the domain of the generalized element by a presentable one.
\end{proof}

Finally, locally presentable topoi provide a natural setting for infinitary extensions of first-order logic. In particular, Kripke--Joyal semantics extends to arbitrary small conjunctions and disjunctions.

\begin{Theorem}[Infinitary Kripke--Joyal semantics]
    \label{thm:fundamental theorem of infinitary KJ semantics}
    Let \(\cC\) be a locally presentable elementary topos, \(\Sigma\) a signature, \(\cM\) a \(\Sigma\)-structure, and \(\alpha\from U \to M_{\vec{x}}\) a generalized element. Then for any small family \((\varphi_i)_{i \in I}\):
    \begin{itemize}
        \item \(\cM \models_\alpha \vec{x}.\bigwedge_I \varphi_i\) iff for all \(i \in I\), \(\cM \models_\alpha \vec{x}.\varphi_i\);
        \item \(\cM \models_\alpha \vec{x}.\bigvee_I \varphi_i\) iff there exists an epimorphic family \((p_i\from V_i \to U)_{i \in I}\) (i.e. the unique morphism \(p\from \coprod V_i \to U\) is an epimorphism) such that for all \(i \in I\), \(\cM \models_{\alpha \circ p_i} \vec{x}.\varphi_i\).
    \end{itemize}
\end{Theorem}

The proof for finite cases from~\cite{MM12} naturally extends to the infinite case. This infinitary extension will be needed for \Los's theorem to handle conjunctions and disjunctions indexed by possibly large families.

\section{\Los's Theorem}
\label{sec:los}

\subsection{Structural Assumptions}
\label{subsec:los:structural-assumptions}

We introduce two structural properties of the models needed for the categorical version of \Los's theorem: internal inhabitedness, and a lifting property along products, corresponding to the axiom of choice in the classical setting.

\subsubsection{Internally Inhabitedness}
\label{subsec:los:structural-assumptions:inhabitedness}

The interpretation of existential formulae in Kripke--Joyal semantics relies on the existence of local witnesses at an object \(X\), that is on \(X\) being internally inhabited\footnote{Note that in a general topos, an object may be internally inhabited without admitting a global element} (i.e., the unique morphism \(!_X \from X \to \terminal\) is an epimorphism). Since the proof of \Los's theorem involves products of structures and existential formulae, it is necessary to control how inhabitedness behaves under products. Thus, we introduce a restricted class of \(\Sigma\)-structures whose sorts are internally inhabited.

\begin{Definition}[Internally inhabited structure]
\label{def:internally inhabited structure}
A \(\Sigma\)-structure \(\cM\) in \(\cC\) is {\bf internally inhabited} if for every sort \(s \in S\), the object \(M_s\) is internally inhabited.
\end{Definition}

This condition is not automatically preserved by products. Indeed, in a general topos, a product of internally inhabited objects may fail to be internally inhabited (see Example~\ref{ex:contre exemple}).
A sufficient condition to prevent inhabitedness from vanishing when considering products is that the terminal object \(\terminal\) is projective, i.e., every epimorphism \(X \to \terminal\) admits a section. This condition guarantees that every internally inhabited object admits a global element, which allows constructing global elements of products by the universal property of products.

\begin{Proposition}
\label{prop: products preserve inhabitedness}
If the terminal object $\terminal$ in \(\cC\) is projective, then for every family $(X_i)_{i \in I}$ of internally inhabited objects in \(\cC\), and every subset $J \subseteq I$, the product $\prod_J X_j$ is internally inhabited.
\end{Proposition}

\begin{proof}
Let $J \subseteq I$ be a set of indices. For every $j \in J$, $X_j$ is internally inhabited, meaning that $!_{X_j}$ is an epimorphism. Since \(\terminal\) is projective, each epimorphism \(!_{X_j}\) admits a section $s_j\from \terminal \to X_j$ such that $!_{X_j} \circ s_j = \id_\terminal$. By the universal property of products, theses sections induce a section $s\from \terminal \to \prod_J X_j$. As $!_{\prod_J X_j} = \prod_J !_{X_j}$, we can deduce that $!_{\prod_J X_j} \circ s = \id_\terminal$ and the product $\prod_J X_j$ is internally inhabited.
\end{proof}

\begin{Corollary}
\label{cor: products preserve inhabitedness}
If the terminal object $\terminal$ in \(\cC\) is projective, then for every family $(\cM_i)_{i \in I}$ of internally inhabited $\Sigma$-structures interpreted \(\cC\), and for every $J \subseteq I$, $\prod_J \cM_j$ is internally inhabited.
\end{Corollary}

The result extends to filtered products in \(\cC\), the extension to \(\Structures{\cC}\) is postponed to Section~\ref{subsec:los:filteredproducts} where the filtered product in \(\Structures{\cC}\) is introduced.

\begin{Proposition}
\label{prop:filtered product is inhabited}
If the terminal object \(\terminal\) of \(\cC\) is projective, then for all families \((X_i)_{i \in I}\) of internally inhabited objects in \(\cC\) and all filters \(F\) on \(I\), the filtered product \(\prod_F X\) is internally inhabited.
\end{Proposition}

\begin{proof}
If \((X_i)_{i \in I}\) is internally inhabited, then \(\prod_I X_i\) is internally inhabited by Proposition~\ref{prop: products preserve inhabitedness} and \(!_{\prod_I X}\) is an epimorphism. Since \(!_{\prod_I X} = {} !_{\prod_F X} \circ \mu_I\), it follows that \(!_{\prod_F X}\) is also an epimorphism.
\end{proof}

Note that the existence of a global element in internally inhabited objects corresponds to a restricted form of the internal choice axiom, sufficient for constructing sections in internally inhabited objects, although it does not imply the full internal axiom of choice.

\subsubsection{Lifting along Products}
\label{subsec:los:structural-assumptions:lifting}

The proof of \Los's theorem for quantified formulae requires lifting generalized elements through canonical product projections. In the classical set-theoretic proof, this step relies on choosing representatives componentwise and therefore uses a form of the axiom of choice. In a general topos, this motivates the following lifting property for families of objects.

\begin{Definition}[Lifting along products]
\label{def:lifting-along-products}
A family \((X_i)_{i \in I}\) of objects of \(\cC\) is {\bf lifting along products} if, for every subset
\(J \subseteq I\), every morphism \(f \from A \to \prod_J X_j\) factors through the canonical projection \(p_{I,J}\).
\end{Definition}

Equivalently, \((X_i)_{i \in I}\) lifting along products if for every object \(A\) in \(\cC\), the mapping
\[
    p_{I,J} \circ \_ \from
    \Hom_\cC(A,\prod_I X_i)
    \to
    \Hom_\cC(A,\prod_J X_j)
\]
is surjective.

\begin{Proposition}
\label{prop:projection-section}
Let \((X_i)_{i \in I}\) be a family of internally inhabited objects of \(\cC\). If the terminal object \(\terminal\) of \(\cC\) is projective, then each projection \(p_{I,J}\), for \(J\subseteq I\), admits a section.
\end{Proposition}

\begin{proof}
Let \(J\subseteq I\). For every \(i \in I \setminus J\), \(X_i\) is internally inhabited, meaning that the unique morphism \(!_{X_i} \from X_i \to \terminal\) is an epimorphism. Since \(\terminal\) is projective, \(!_{X_i}\) admits a section \(s_i \from \terminal \to X_i\) such that \(\id_\terminal ={} !_{X_i} \circ s_i\).

We now construct a section \(s_{I,J} \from \prod_J X_j \to \prod_I X_i\) for \(p_{I,J}\):
\begin{itemize}
    \item For each \(j \in J\), define the \(j\)-th component of \(s_{I,J}\) as the projection \(p_{J,j}\).
    \item For each \(i \in I \setminus J\), define the \(i\)-th component of \(s_{I,J}\) as the morphism \(s_i \circ{} !_{\prod_J X_j}\).
\end{itemize}
By construction, for every \(j \in J\), \(p_{I,j} \circ s_{I,J} = p_{J,j}\). Thus \(p_{I,J} \circ s_{I,J} = \id_{\prod_J X_j}\).
\end{proof}

Projectivity of the terminal object and internal inhabitedness ensure that, for every \(J\subseteq I\), the canonical projection \(p_{I,J}\) admits a section which entails this lifting property.

\begin{Corollary}
\label{cor:lifting-from-projective-terminal}
Under the condition that the terminal object \(\terminal\) of \(\cC\) is projective, any family of internally inhabited objects of \(\cC\) is lifting along products.
\end{Corollary}

\begin{proof}
Since \(p_{I,J}\) admits a section \(s_{I,J}\) by Proposition~\ref{prop:projection-section}, any morphism \(f \from X \to \prod_J X_j\) satisfies
\[
    f = p_{I,J} \circ (s_{I,J} \circ f).
\]
Hence \((X_i)_{i \in I}\) is lifting along products.
\end{proof}

Families that are lifting along product ensure that canonical projections between subproducts become epimorphisms.

\begin{Proposition}
\label{prop:subproduct-projections-epi}
Let \((X_i)_{i \in I}\) be a lifting along product family of objects of \(\cC\). Then each canonical projection \(p_{I,J}\), for \(J\subseteq I\), is an epimorphism.
\end{Proposition}

\begin{proof}
Let \(f_1,f_2\from \prod_J X_j \to X\) be morphisms such that \(f_1 \circ p_{I,J} = f_2 \circ p_{I,J}\). Let \(g\from A \to \prod_J X_j\) be a morphism. Since \((X_i)_{i \in I}\) is lifting along product, there exists \(h\from A \to \prod_I X_i\) such that \(g = p_{I,J} \circ h\). Therefore,
\begin{align*}
f_1 \circ g
& = f_1 \circ p_{I,J} \circ h \\
& = f_2 \circ p_{I,J} \circ h \\
& = f_2 \circ g
\end{align*}
or diagrammatically
\[\begin{tikzcd}[column sep=scriptsize]
    {\prod_I X_i} & {\prod_J X_j} & X \\
    A
    \arrow["{p_{I,J}}", from=1-1, to=1-2]
    \arrow["{f_1}", shift left, from=1-2, to=1-3]
    \arrow["{f_2}"', shift right, from=1-2, to=1-3]
    \arrow["h", from=2-1, to=1-1]
    \arrow["g"', from=2-1, to=1-2]
\end{tikzcd}\]
and then, \(f_1 \circ g = f_2 \circ g\), for every morphism \(g\from A \to \prod_J X_i\). Thus, \(f_1 = f_2\) and \(p_{I,J}\) is an epimorphism.
\end{proof}

In locally presentable categories, lifting along products naturally extends to filtered product when considering morphisms from presentable objects.

\begin{Proposition}
\label{prop:lifting-filtered-products}
Let \(\cC\) be a locally presentable category with arbitrary products and filtered products. Let \((X_i)_{i \in I}\) be a lifting along product family of objects of \(\cC\). Then, every morphism \(f \from A \to \prod_F X\) factors through \(\mu_I\) whenever \(A\) is a presentable object of \(\cC\).
\end{Proposition}

Similar to Definition~\ref{def:lifting-along-products}, the factorisation condition is equivalent to: for every presentable object \(A\) of \(\cC\), the mapping
\[
    \mu_I \circ \_
    \from
    \Hom_\cC(A,\prod_I X_i)
    \to
    \Hom_\cC(A,\prod_F X)
\]
is surjective.

\begin{proof}
Let \(f \from A \to \prod_F X\) be a morphism with \(X\) presentable. Since filtered products are filtered colimits of subproducts and \(A\) is presentable, there exists \(J \in F\) and a morphism \(g \from A \to \prod_J X_j \) such that \(f = \mu_J \circ g\). Since \((X_i)_{i \in I}\) is lifting along products, there exists \(h \from A \to \prod_I X_i\)
such that \(g = p_{I,J} \circ h\). Therefore, 
\begin{align*}
f   &= \mu_J \circ g \\
    &= \mu_J \circ p_{I,J} \circ h\\
    &= \mu_I \circ h
\end{align*}
or diagrammatically
\[\begin{tikzcd}[column sep=scriptsize]
	{\prod_I X_i} & {\prod_J X_j} & {\prod_F X} \\
	& A
	\arrow["{p_{I,J}}", from=1-1, to=1-2]
	\arrow["{\mu_I}", curve={height=-30pt}, from=1-1, to=1-3]
	\arrow["{\mu_J}", from=1-2, to=1-3]
	\arrow["h", from=2-2, to=1-1]
	\arrow["g"', from=2-2, to=1-2]
	\arrow["f"', from=2-2, to=1-3]
\end{tikzcd}\]
and then, every morphism from a presentable object to \(\prod_F X\) factors through \(\mu_I\).
\end{proof}

\subsection{Filtered Products in \mathinhead{\Structures{\cC}}{Sigma-Structures}}
\label{subsec:los:filteredproducts}

Let \(\cC\) be a locally presentable elementary topos, \(I\) a set, \(F\) a filter on \(I\), and \((\cM_i)_{i\in I}\) a family of \(\Sigma\)-structures in \(\cC\).

We first define a candidate structure \(\cM_F\) and then show that it is indeed a \(\Sigma\)-structure. Define the sortwise filtered-product construction \(\cM_F\) as follows.
\begin{itemize}
    \item for each \(s \in S\), \((M_F)_s = \prod_F (M_i)_s\) is the filtered product of \(((M_i)_s)_{i \in I}\) modulo \(F\). For \(J \in F\), we thus have structure maps \((\alpha_J)_s \from \prod_J (M_j)_s \to (M_F)_s\) for each sort \(s \in S\). 

    \item for each function symbol \(f\from s_1 \times \ldots \times s_n \to s\), the morphisms \((\alpha_J)_s \circ f^{\prod_J\cM_j} \from \prod_J(M_j)_{s_1}\times\cdots\times \prod_J(M_j)_{s_n} \to (M_F)_s\) form a cocone. Since filtered colimits commute with finite products in \(\cC\), they induce a unique morphism \(f^{\cM_F} \from (M_F)_{s_1} \times \ldots \times (M_F)_{s_n} \to (M_F)_{s}\).
    
    \item for each relation symbol \(r\from s_1 \times \ldots \times s_n\), \(r^{\cM_F}\) is the filtered colimit of the diagram \(J \mapsto \prod_J r^{\cM_J}\) for \(J \in F\), where \(r^{\prod_J\cM_j}\) is the interpretation of \(r\) in the product structure \(\prod_J\cM_j\).
\end{itemize}

\begin{Proposition}
\label{prop:sortwise-filtered-product-is-structure}
The sortwise filtered-product construction \(\cM_F\) defines a \(\Sigma\)-structure.
\end{Proposition}

\begin{proof}
The interpretations of function symbols are well-defined because filtered colimits commute with finite products in \(\cC\) (Lemma~\ref{lma:filtered-colimits-commutes-small-limits}).

For a relation symbol \(r:s_1\times\cdots\times s_n\), the interpretation \(r^{\cM_F}\) is obtained as the filtered colimit of the subobjects \(r^{\cM_i}\). Since filtered colimits commute with finite limits in \(\cC\) (Lemma~\ref{lma:filtered-colimits-commutes-small-limits}), they preserve monomorphisms. Hence \(r^{\cM_F}\) is a subobject of \((M_F)_{s_1} \times \ldots \times (M_F)_{s_n}\).

Therefore all function and relation symbols are interpreted correctly, and \(\cM_F\) is indeed a \(\Sigma\)-structure.
\end{proof}

\begin{Proposition}
\label{prop:sortwise-filtered-product-gives-morphisms}
For each \(J \in F\), the families of morphisms \(((\alpha_J)_s \from \prod_J (M_j)_s \to (M_F)_s)_{s \in S}\) defines a \(\Sigma\)-structure morphism \(\alpha_J \from \prod_J \cM_j \to \cM_F\).
\end{Proposition}

\begin{proof}
For function symbols \(f\from s_1 \times \ldots \times s_n \to s\), the expected diagrams commute by construction of \(f^{\cM_F} \from (M_F)_{s_1} \times \ldots \times (M_F)_{s_n} \to (M_F)_{s}\) as the unique morphisms induced by the commutativity of filtered colimits with finite products in \(\cC\). For relation symbols \(r\from s_1 \times \ldots \times s_n\), the existence of the morphisms between the codomain of the subobjects follows from the construction of \(r^{\cM_F}\) as a colimit.
\end{proof}

\begin{Proposition}
\label{prop:sortwise-filtered-product-is-the-filtered-product}
For every family $(\cM_i)_{i \in I}$ of $\Sigma$-structures in \(\cC\), the sortwise filtered-product construction \(\cM_F\) is the filtered product of \((\cM_i)_{i \in I}\) modulo $F$.
\end{Proposition}

\begin{proof}
    By Propositions~\ref{prop:sortwise-filtered-product-is-structure} and~\ref{prop:sortwise-filtered-product-gives-morphisms}, the family \(\alpha = (\alpha_J)_{J \in F}\) forms a cocone \(A_F \Rightarrow \cM_F\) where the functor \(A_F \from F \to \Structures{\cC}\) maps indices \(J\) of the filter \(F\) to models \(\prod_J \cM_j\) and inclusions \(J \subseteq J'\) to projections \(p_{J',J}\).
    
    Let \(\nu = (\nu_J \from \prod_J \cM_j \to \cN)_{J \in F}\) be another cocone and let us show that there exists a unique \(\Sigma\)-structure morphism \(\theta\from \cM_F \to \cN\) such that \(\theta \circ \alpha_J = \nu_J\) for every \(J \in F\).
    Since \((M_F)_s\) is the filtered product of \(((M_i)_s)_{i \in I}\) modulo \(F\) for each sort \(s \in S\), there exists a unique morphism \(\theta_s\from (M_F)_s \to N_s\) such that \(\theta_s \circ (\alpha_J)_S = (\nu_J)_s\) for each sort \(s\) in \(S\) and we have a family \(\theta = (\theta_s)_{s \in S}\) of morphisms in \(\cC\). It remains to prove that \(\theta\) is a morphism in \(\Structures{\cC}\).

    Let \(f\from s_1 \times \ldots \times s_n \to s\) be a function symbol. We aim to prove that \(\theta_s \circ f^{\cM_F} = f^{\cN} \circ (\theta_{s_1}, \ldots, \theta_{s_n})\), which amounts to proving that the diagram
    \[\begin{tikzcd}[column sep=huge,row sep=3.15em]
        {(\prod_J M_j)_{s_1} \times \ldots \times (\prod_J M_j)_{s_n}} & {(\prod_J M_j)_{s}} \\
        {(M_F)_{s_1} \times \ldots \times (M_F)_{s_n}} & {(M_F)_s} \\
        {N_{s_1} \times \ldots \times N_{s_n}} & {N_s}
        \arrow["{f^{\prod_J \cM_j}}", from=1-1, to=1-2]
        \arrow["{(\alpha_J)_{s_1},\ldots, (\alpha_J)_{s_n}}", from=1-1, to=2-1]
        \arrow["{(\nu_J)_{s_1}, \ldots, (\nu_J)_{s_n}}"', shift right=5, curve={height=50pt}, from=1-1, to=3-1]
        \arrow["{(\alpha_J)_s}", from=1-2, to=2-2]
        \arrow["{(\nu_J)_s}", shift left=2, curve={height=-30pt}, from=1-2, to=3-2]
        \arrow["{f^{\cM_F}}", from=2-1, to=2-2]
        \arrow["{\theta_{s_1},\ldots, \theta_{s_n}}", from=2-1, to=3-1]
        \arrow["{\theta_s}", from=2-2, to=3-2]
        \arrow["{f^\cN}"', from=3-1, to=3-2]
    \end{tikzcd}\]
    commutes for every \(J\in F\). Let \(J\in F\) be a set of indices. By construction of \(f^{\cM_F}\), 
    \[
        f^{\cM_F} \circ \bigl( (\alpha_J)_{s_1}, \ldots, (\alpha_J)_{s_n} \bigr) = (\alpha_J)_{s} \circ f^{\prod_J \cM_j},
    \]
    i.e., the top square of the diagram commutes. Besides, the outer square also commutes since \(\nu_J\) is a morphism in \(\Structures{\cC}\), i.e.,
    \[
        (\nu_J)_s \circ f^{\prod_J \cM_j} = f^\cN \circ \bigl( (\nu_J)_{s_1}, \ldots, (\nu_J)_{s_n} \bigr).
    \]
    Thus,
    \begin{align*}
        \theta_s \circ f^{\cM_F} \circ \bigl( (\alpha_J)_{s_1}, \ldots, (\alpha_J)_{s_n} \bigr)
        &= \theta_s \circ (\alpha_J)_{s} \circ f^{\prod_J \cM_j}\\
        &= (\nu_J)_s \circ f^{\prod_J \cM_j}\\
        &= f^\cN \circ \bigl( (\nu_J)_{s_1}, \ldots, (\nu_J)_{s_n} \bigr)\\
        &= f^\cN \circ (\theta_{s_1},\ldots, \theta_{s_n}) \circ \bigl( (\alpha_J)_{s_1}, \ldots, (\alpha_J)_{s_n} \bigr).
    \end{align*}
    Besides, for every sort \(s\in S\), the family \(((\alpha_J)_s)_{J \in F}\) is jointly epic in \(\cC\) (as a colimit cocone). Since filtered colimits commute with finite products (Lemma~\ref{lma:filtered-colimits-commutes-small-limits}), it follows that \(((\alpha_J)_{s_1}, \ldots, (\alpha_J)_{s_n})_{J \in F}\) is a jointly epic family. Since equality of morphisms can be tested after precomposition with a jointly epic family, the claim follows: \(\theta_s \circ f^{\cM_F} = f^{\cN} \circ (\theta_{s_1}, \ldots, \theta_{s_n})\).

    Let \(r\from s_1 \times \ldots \times s_n\) be a relation symbol. For every \(J\in F\), \(\nu_J\) is a morphism in \(\Structures{\cC}\) so it preserves \(r\) and there exists a morphism \(O_{\prod_J\cM_j}\to O_\cN\) such that the following diagram commutes.
    \[\begin{tikzcd}[column sep=huge]
        {O_{\prod_J \cM_j}} & {\prod_J (M_j)_{s_1} \times \ldots \times \prod_J (M_j)_{s_n}} \\
        {O_\cN} & {N_{s_1} \times \ldots \times N_{s_n}}
        \arrow["{r^{\prod_J \cM_j}}", tail, from=1-1, to=1-2]
        \arrow[from=1-1, to=2-1]
        \arrow["{(\nu_J)_{s_1}, \ldots, (\nu_J)_{s_n}}", from=1-2, to=2-2]
        \arrow["{r^{\cN}}"', tail, from=2-1, to=2-2]
    \end{tikzcd}\]
    The family of left vertical morphisms \((O_{\prod_J \cM_j} \to O_\cN)_{J \in F}\) is compatible with the filtered diagram defining \(r^{\cM_F} = \prod_F(r^{\cM_i})\). Let \(O_{\cM_F}\) be the domain of
    \(r^{\cM_F}\).
    The universal property of colimits yields a unique morphism \(O_{\cM_F} \to O_\cN\) such that the diagram
    \[\begin{tikzcd}[column sep=huge,row sep=3.15em]
        {O_{\prod_J \cM_j}} & {\prod_J (M_j)_{s_1} \times \ldots \times \prod_J (M_j)_{s_n}} \\
        {O_{\cM_F}} & {(M_F)_{s_1} \times \ldots \times (M_F)_{s_n}} \\
        {O_\cN} & {N_{s_1} \times \ldots \times N_{s_n}}
        \arrow["{r^{\prod_J \cM_j}}", tail, from=1-1, to=1-2]
        \arrow[from=1-1, to=2-1]
        \arrow["{(\nu_J)_{s_1}, \ldots, (\nu_J)_{s_n}}"', shift right=2, curve={height=24pt}, from=1-1, to=3-1]
        \arrow["{(\alpha_J)_{s_1},\ldots, (\alpha_J)_{s_n}}"', from=1-2, to=2-2]
        \arrow["{(\nu_J)_{s_1}, \ldots, (\nu_J)_{s_n}}", shift left=5, curve={height=-50pt}, from=1-2, to=3-2]
        \arrow["{r^{\cM_F}}", tail, from=2-1, to=2-2]
        \arrow[from=2-1, to=3-1]
        \arrow["{\theta_{s_1},\ldots, \theta_{s_n}}"', from=2-2, to=3-2]
        \arrow["{r^\cN}"', tail, from=3-1, to=3-2]
    \end{tikzcd}\]
    commutes (for every \(J \in F\)).

    Thereafter, \(\theta\) is a morphism in \(\Structures{\cC}\) and it remains to prove that \(\theta\) is unique. Let \(\theta'\from \cM_F\to\cN\) be a morphism in \(\Structures{\cC}\) such that \(\theta' \circ \alpha_J = \nu_J\) for every \(J \in F\). Then, for every sort \(s\), \(\theta'_s \circ (\alpha_J)_s = (\nu_J)_s = \theta_s \circ (\alpha_J)_s\). Since \((M_F)_s\) is the colimit of the underlying diagram for the sort \(s\), the universal property of the colimit implies \(\theta'_s=\theta_s\). Hence, we conclude that \(\theta' = \theta\).
\end{proof}

In the following, we will no longer talk about the sortwise filtered-product construction but about the filtered product of \(\Sigma\) structures and, therefore, come back to the notation \(\prod_F \cM\) with colimit morphisms \(\mu_J\from \prod_J \cM_j  \to \prod_F \cM\). Since we established that the construction yields the filtered product, we can now provide some key technical ingredients for the proof of \Los's theorem. First, we show that the interpretation of terms is compatible with filtered products.

\begin{Proposition}
\label{prop:terms-commute-filtered-products}
For every term-in-context \(\vec{x}.t\) with \(t:s\) and \(\vec{x} = (x_1:s_1,\ldots,x_n:s_n)\), the interpretation of \(t\) commutes with filtered products. More precisely, for every \(J \in F\), the diagram
\[\begin{tikzcd}[column sep=huge]
	{(\prod_J M_j)_{\vec{x}}} & {(\prod_J M_j)_{s}} \\
	{(\prod_F M)_{\vec{x}}} & {(\prod_F M)_{s}}
	\arrow["{\sem[\prod_J \cM_j]{\vec{x}.t}}", from=1-1, to=1-2]
	\arrow["{(\mu_J)_{\vec{x}}}"', from=1-1, to=2-1]
	\arrow["{(\mu_J)_s}", from=1-2, to=2-2]
	\arrow["{\sem[\prod_F \cM]{\vec{x}.t}}"', from=2-1, to=2-2]
\end{tikzcd}\]
commutes.
\end{Proposition}

\begin{proof}
We proceed by induction on the structure of the term \(t\) and consider some \(J \in F\).

If \(t\) is a variable, then it corresponds to an \(x_k\) in \(\vec{x}\) and \(\sem[\cM]{\vec{x}.x_k}\) is the corresponding projection.
Since \((\prod_F M)_{\vec{x}}\) is obtained sortwise as a filtered colimit and filtered colimits commute with finite products, the morphism \((\mu_J)_{\vec{x}}\) is the product of the morphisms
\((\mu_J)_{s_1},\ldots,(\mu_J)_{s_n}\). Therefore, the result follows from the universal property of products.

If \(t\) is \(f(t_1, \ldots t_m)\) for some function symbol \(f\from s'_1\times\cdots\times s'_m\to s\) and \(t_1\from s'_1, \ldots, t_m\from s'_m\) such that the statement holds for \(t_1,\ldots, t_m\). Recall that, by definition,
\[
    \sem[\prod_F \cM]{\vec{x}.t} = f^{\prod_F \cM} \circ \bigl( \sem[\prod_F \cM]{\vec{x}.t_1}, \ldots, \sem[\prod_F \cM]{\vec{x}.t_m}
\bigr).
\]

Now, the following diagram commutes
\begin{center}
\vspace{-1em}
\begin{tikzpicture}[baseline=(current bounding box.center), transform shape, scale=.9]
\node[inner sep=0, outer sep=0]{
\begin{tikzcd}[sep=3em]
	{(\prod_J M_j)_{\vec{x}}} &&&& {(\prod_J M_j)_{s'_1} \times \ldots \times (\prod_J M_j)_{s'_m}} & {(\prod_J M_j)_{s}} \\
	{(\prod_F M)_{\vec{x}}} &&&& {(\prod_F M)_{s'_1} \times \ldots \times (\prod_F M)_{s'_m}} & {(\prod_F M)_{s}}
	\arrow["{\sem[{\prod_J} \cM_j]{\vec{x}.t_1}, \ldots, \sem[{\prod_J} \cM_j]{\vec{x}.t_m}}", from=1-1, to=1-5]
	\arrow["{(\mu_J)_{\vec{x}}}"', from=1-1, to=2-1]
	\arrow["{f^{{\prod_J} \cM_j}}", from=1-5, to=1-6]
	\arrow["{(\mu_J)_{s'_1},\ldots,(\mu_J)_{s'_m}}"', from=1-5, to=2-5]
	\arrow["{(\mu_J)_s}", from=1-6, to=2-6]
	\arrow["{\sem[{\prod_F} \cM]{\vec{x}.t_1}, \ldots, \sem[{\prod_F} \cM]{\vec{x}.t_m}}"', from=2-1, to=2-5]
	\arrow["{f^{{\prod_F} \cM}}"', from=2-5, to=2-6]
\end{tikzcd}
};
\end{tikzpicture}
\end{center} 
where the commutativity of the left square follows from the induction hypothesis together with the universal property of products, and that of the right square from the definition of \(f^{\prod_F \cM}\), concluding the proof.
\end{proof}

An immediate consequence of the previous result is that the interpretation of atomic formulae also commutes with filtered products.

\begin{Corollary}
\label{cor:atomic-commute-filtered-products}
For every atomic formula-in-context \(\vec{x}.r(\vec{t})\) with \(r\from s'_1 \times \ldots \times s'_m\), \(\vec{t} = (t_1\from s'_1, \ldots, t_m\from s'_m)\); and \(\vec{x} = (x_1\from s_1, \ldots, x_n\from s_n)\), the interpretation of \(r(\vec{t})\) commutes with filtered products. More precisely, the monomorphism \(\sem[\prod_F \cM]{\vec{x}.r(\vec{t})}\) is the filtered product of the family of monomorphisms \(\bigl(\sem[\cM_i]{\vec{x}.r(\vec{t})} \bigr)_{i\in I}\). Equivalently \(\set{\vec{x}.r(\vec{t})}^{\prod_F \cM} = \prod_F \set{\vec{x}.r(\vec{t})}^\cM\).
\end{Corollary}

\begin{proof}
By definition, \(\sem[\prod_F \cM]{\vec{x}.r(\vec{t})} = r^{\prod_F \cM} \circ \sem[\prod_F \cM]{\vec{x}.\vec{t}}\) and \(\prod_F \sem[\cM]{\vec{x}.r(\vec{t})} = \prod_F(r^\cM) \circ \prod_F \sem[\cM]{\vec{x}.\vec{t}}\). The result then follows from Proposition~\ref{prop:terms-commute-filtered-products} and the definition of \(r^{\prod_F \cM}\) as the filtered product \(\prod_F(r^\cM)\).
\end{proof}

Second, we show that filtered products preserve internal inhabitedness.

\begin{Proposition}
\label{prop:sortwise-filtered-product-is-inhabited}
If the terminal object \(\terminal\) of \(\cC\) is projective, then for all families $(\cM_i)_{i \in I}$ of internally inhabited $\Sigma$-structures in \(\cC\), the filtered product \(\prod_F \cM\) is internally inhabited.
\end{Proposition}

\begin{proof}
By Proposition~\ref{prop:filtered product is inhabited}, each sort of \(\prod_F \cM\) is internally inhabited. Hence \(\prod_F \cM\) is internally inhabited.
\end{proof}

Projectivity of the terminal object ensures that internal inhabitedness is preserved by filtered product, which does not hold on its own, as illustrated by the following example, showing that \Los's theorem does not hold without this condition.\footnote{This counterexample is adapted from one suggested by a reviewer of a previous version of this paper, whom we gratefully acknowledge.}

\begin{Example}
\label{ex:contre exemple}
Let us consider the topological space $X$ defined by the set of positive integers $\bN$ equipped with the cofinite topology \(\cT\), that is
\[
\cT = \set{O \subseteq \bN \mid O = \emptyset \text{ or } \bN \setminus O \text{ is finite.}}
\]
In this topology, every open set is compact. Thus, the Grothendieck topology on the poset of open set \(\cO(X)\) is generated by finite covering families. It follows that the category \(Sh(X)\) of sheaves over \(X\) is locally finitely presentable. For every $n \in \bN$, let $U_n = \bN \setminus \{n\}$ and consider the sheaf 
\[
\Function{F_n}{\cO(X)^{op}}{\Set}{V}{\Hom_{\cO(X)}(V,U_n)}.
\]
Thus, for every open \(V\),
\[
F_n(V) = 
\left\{
\begin{array}{cl}
\terminal & \mbox{if $V \subseteq U_n$} \\
\emptyset & \mbox{otherwise}
\end{array}
\right.
\]

Let $\Sigma$ be the signature consisting of a single sort, so that a $\Sigma$-structure in $Sh(X)$ is just an object of $Sh(X)$.
We consider \(I = \bN\), and \(F\) a non-principal ultrafilter on \(\bN\). In particular, any \(J \in F\) is infinite. Now, for every $n \in \bN$, we define the $\Sigma$-structure $\cM_n = F_{2n} + F_{2n+1}$, that is the coproduct of the sheaves $F_{2n}$ and $F_{2n+1}$.

We first observe that every \(\cM_n\) is internally inhabited. Indeed, \(U_{2n}\) and \(U_{2n+1}\) form an open cover of \(\bN\), meaning that, for each $n \in \bN$, $\cM_n \Rightarrow \terminal$ is an epimorphism.
On the other hand, there is no global section $s\from \terminal \Rightarrow \cM_n$. Such a global section would map any open set $V$ to an element in $\cM_n(V)$, yet
\[
\cM_n(\bN)
=
F_{2n}(\bN)+F_{2n+1}(\bN)
=
\emptyset,
\]
since \(\bN\not\subseteq U_{2n}\) and \(\bN\not\subseteq U_{2n+1}\).
In particular, the terminal object is not projective in \(Sh(X)\).

Now, let \(J \in F\). For every non-empty open \(V \in \cO(X)\), there are only finitely many \(n \in \bN\) such that \(V\subseteq U_{2n}\) or \(V\subseteq U_{2n+1}\). Thus, \(\prod_J \cM_j(V) = \emptyset\) (since \(J\) is infinite). It follows that \(\prod_{j\in J}\cM_j\simeq\initial\) is trivial, and then \(\prod_F \cM \simeq \initial\), as a filtered colimit of empty sheaves. In particular, although every factor \(\cM_n\) is internally inhabited, their filtered product is not.

\end{Example}

This example gives precisely a counterexample that motivates the requiring the projectivity of \(\terminal\). Recall that in Kripke--Joyal semantics, a structure \(\cM\) satisfies \([].(\exists x\from s.\top)\) (for a sort \(s\)) if and only if the unique morphism \(M_s \to \terminal\) is an epimorphism, which holds for every \(\cM_n\) by internally inhabitedness. Thus
\[
\set{n\in \bN\mid\cM_n\models[].(\exists x\from s.\top)}=\bN\in F.
\]
However, \(\prod_F\cM\simeq\initial\), so its unique morphism to \(\terminal\) is not an epimorphism, meaning that
\[
\prod_F\cM\not\models\exists x.\top.
\]

In particular,
\[
\set{i\in I\mid\cM_i\models\varphi}\in F \; \text{implies} \; \prod_F\cM\models\varphi
\]
fails in \(Sh(X)\) for the formula \(\exists x.\top\).

The hypothesis that \(\terminal\) is projective rules out this phenomenon: by Proposition~\ref{prop:sortwise-filtered-product-is-inhabited}, filtered products of internally inhabited structures are then internally inhabited.

\subsection{Fundamental Theorem}
\label{subsec:los's theorem}

We can now establish the main technical result underlying our generalization of \Los's theorem for first-order structure in elementary topoi for Kripke--Joyal semantics. In the set-theoretic setting, \Los's theorem is proved by induction on the formulae. We follow the same strategy, but the induction behaves differently in the two directions of the equivalence. In particular, the two implications do not in general hold under the
same hypotheses.

We therefore separate them into two notions. Adapting terminology from Diaconescu's treatment of filtered products in institution theory~\cite[Chap. 6]{Dia08}, factor stability expresses the implication from satisfaction in the filtered product to satisfaction on a set of factors, while product stability expresses the converse implication. Thus, the core part of this subsection is to establish closure properties for these two classes of formulae, providing the logical fragments for which each direction of \Los's theorem can be obtained, and making explicit where the additional assumptions on the topos enter. 

Throughout this subsection, let \(\cC\) be a locally presentable topos, \(I\) be a set, \(F\) be a filter on \(I\), and \((\cM_i)_{i\in I}\) be a family of \(\Sigma\)-structures in \(\cC\).

\begin{Notation}
For a formula-in-context \(\vec{x}.\varphi\) and a generalized element \(\alpha \from U \to (\prod_I M_i)_{\vec{x}}\), we write \(\SatisIndexSet{\vec{x}.\varphi}\) for the set of indices \(i\) such that \(\cM_i\) satisfies \(\vec{x}.\varphi\) at the generalized element \((p_{I,i})_{\vec{x}} \circ \alpha\), i.e., the satisfaction index set
\[
\SatisIndexSet{\vec{x}.\varphi} = \set{i \in I \mid \cM_i \models_{(p_{I,i})_{\vec{x}} \circ \alpha} \vec{x}.\varphi}.
\]
\end{Notation}

We separate the two directions of \Los's theorem and introduce the following two notions (understood relative to the fixed filter \(F\), and the family \((\cM_i)_{i\in I}\)).

\begin{Definition}[Stability under factors and products]
Let \(\alpha \from U \to (\prod_I M_i)_{\vec{x}}\) be a generalized element. A formula-in-context \(\vec{x}.\varphi\) is {\bf stable under filtered factor at \(\alpha\)}, or \(\alpha\)-factor-stable for short, if
\[
    \prod_F \cM \models_{(\mu_I)_{\vec{x}} \circ \alpha} \vec{x}.\varphi \; \text{implies} \; \SatisIndexSet{\vec{x}.\varphi} \in F.
\]
The formula is {\bf stable under filtered factor}, or factor-stable, if it is so at every generalized element.

A formula-in-context \(\vec{x}.\varphi\) is {\bf stable under filtered product at \(\alpha\)}, or \(\alpha\)-product-stable for short, if
\[
    \SatisIndexSet{\vec{x}.\varphi} \in F \; \text{implies} \; \prod_F \cM \models_{(\mu_I)_{\vec{x}} \circ \alpha} \vec{x}.\varphi.
\]
The formula is {\bf stable under filtered product}, or product-stable, if it is so at every generalized element.
\end{Definition}

We use the following notation to distinguish the scope over which stability is required. The subscript \(\alpha\) refers to a fixed generalized element, while \(\mathrm{pr}\) and \(\mathrm{glb}\) restrict the domains of the generalized elements to presentable objects and to the terminal object, respectively.

\begin{Notation}
We write \(\cF_\alpha\) for the class of formulae stable under filtered factor at \(\alpha\),
\(
\cF = \bigcap_{\alpha} \cF_\alpha,
\)
for the class of formulae stable under filtered factor,
\[
\cF_{\mathrm{pr}} = \bigcap_{\substack{\alpha \from U \to (\prod_I M_i)_{\vec{x}}\\ U \text{ presentable}}} \cF_\alpha,
\]
for the class of formulae stable under filtered factor at every generalized element whose domain is presentable, and
\[
\cF_{\mathrm{glb}} = \bigcap_{\alpha \from \terminal \to (\prod_I M_i)_{\vec{x}}} \cF_\alpha,
\]
for the class of formulae stable under filtered factor at every global element.

Similarly, we write \(\cP_\alpha\), \(\cP\), \(\cP_{\mathrm{pr}}\), and \(\cP_{\mathrm{glb}}\) for formulae stable under filtered product at the associated generalized elements.
\end{Notation}

Note that the classes \(\cF_\alpha\), \(\cP_\alpha\), \(\cF\), \(\cP\), \(\cF_{\mathrm{pr}}\), \(\cP_{\mathrm{pr}}\), \(\cF_{\mathrm{glb}}\), and \(\cP_{\mathrm{glb}}\) are all understood relative to the fixed filter \(F\) and family \((\cM_i)_{i\in I}\). Additionally, when \(\terminal\) is presentable \(\cF_{\mathrm{pr}} \subseteq \cF_{\mathrm{glb}}\) and \(\cP_{\mathrm{pr}} \subseteq \cP_{\mathrm{glb}}\).

The result now proceeds by establishing closure properties of these classes. The first group of properties is purely structural and provides the inductive basis for positive connectives. The ultrafilter assumption is then used for the classical Boolean steps, while projectivity of \(\terminal\) allows local witnesses to be replaced by global ones. Finally, two-valuedness is used for the remaining negation argument and extends the closure of disjunction.

\begin{Theorem}[\Los{} closure properties]
\label{thm:Los:kjsem}
Let \((\cM_i)_{i \in I}\) be a family of internally inhabited \(\Sigma\)-structures in \(\cC\) and \(\alpha \from U \to (\prod_I M_i)_{\vec{x}}\) be a generalized element.
\begin{enumerate}
\item\label{item:atomic-factors} Atomic formulae belong to \(\cF_{\mathrm{pr}}\).

\item\label{item:atomic-products} Atomic formulae belong to \(\cP\).

\item\label{item:conjunction} \(\cF_\alpha\) and \(\cP_\alpha\) are closed under finite conjunctions.

\item\label{item:infinite-conjunction} \(\cP_\alpha\) is closed under arbitrary conjunctions.

\item\label{item:disjunction-filters} \(\cF\) is closed under finite disjunctions.
\end{enumerate}
If \(F\) is an ultrafilter, then:
\begin{enumerate}[resume]
\item\label{item:implication} If \(\varphi \in \cF\) and \(\psi \in \cP\) then \((\varphi \Rightarrow \psi) \in \cP\).

\item\label{item:negation-factor-to-product} If \(\varphi \in \cF\), then \((\neg \varphi)\in \cP\).
\end{enumerate}
If the terminal object \(\terminal\) is projective, then:
\begin{enumerate}[resume]

\item\label{item:existential-quantification-factors} \(\cF_{\mathrm{pr}}\) is closed under existential quantification.

\item\label{item:existential-quantification-products} \(\cP_{\mathrm{glb}}\,\) is closed under existential quantifier.
\end{enumerate}
If \(F\) is an ultrafilter and \(\terminal\) is projective, then:
\begin{enumerate}[resume]     
\item\label{item:universal-quantification-factor} \(\cF\) is closed under universal quantification.

\item\label{item:universal-quantification-product} \(\cP\) is closed under universal quantification.
\end{enumerate}
If \(F\) is an ultrafilter, \(\terminal\) is projective, and \(\cC\) is two-valued, then:
\begin{enumerate}[resume]
\item\label{item:disjunction-products} $\cP_{\alpha}$ is closed under finite disjunction, whenever \(\alpha\) is a global element.

\item\label{item:negation-factor} $\cF_{\alpha}$ is closed under negation, whenever \(\alpha\) is a global element.
\end{enumerate}
\end{Theorem}

The proof is organized according to the preceding closure properties.
Its main technical aspects stem from the need to reason locally as per Kripke--Joyal satisfaction, i.e., per generalized element \(\alpha\), with witnesses of (un)satisfaction built via universal properties. 

\begin{proof}
We prove each item seperately. Items~\ref{item:atomic-factors} and~\ref{item:atomic-products} establish the case of atomic formulae. Items~\ref{item:conjunction}, \ref{item:infinite-conjunction}, \ref{item:disjunction-filters}, \ref{item:negation-factor-to-product}, \ref{item:disjunction-products}, and \ref{item:negation-factor} correspond to propositional connectives. Items~\ref{item:existential-quantification-factors}, \ref{item:existential-quantification-products}, \ref{item:universal-quantification-factor}, and~\ref{item:universal-quantification-product} treat the quantifiers.
Let \((\cM_i)_{i \in I}\) be a family of internally inhabited \(\Sigma\)-structures in \(\cC\) and \(\alpha \from U \to (\prod_I M_i)_{\vec{x}}\) be a generalized element, \(\varphi\), \(\psi\) and \((\varphi_k)_{k \in K}\) be formulae.

\begin{enumerate}

\item[\ref{item:atomic-factors}.]
Suppose that \(\prod_F \cM \models_{(\mu_I)_{\vec{x}} \circ \alpha} \vec{x}.r(\vec{t})\). We show that the satisfaction index set \(\SatisIndexSet{\vec{x}.r(\vec{t})}\) belongs to \(F\). By assumption, there exists a morphism \(m\from U \to \set{\vec{x}\mid r(\vec{t})(\vec{x})}_{\prod_F\cM}\) such that
\[
\sem[\prod_F\cM]{\vec{x}.r(\vec{t})}\circ m = (\mu_I)_{\vec{x}}\circ \alpha.
\]
By Corollary~\ref{cor:atomic-commute-filtered-products}, the object \(\set{\vec{x}\mid r(\vec{t})(\vec{x})}_{\prod_F\cM} \) is the filtered product of the family \((\set{\vec{x}\mid r(\vec{t})(\vec{x})}_{\cM_i})_{i\in I}\). Since \(U\) is presentable, the morphism \(m\) factors through one of the colimit morphisms. Hence there exist \(J\in F\) and a morphism \(\delta_J\from U \to \set{\vec{x}\mid r(\vec{t})(\vec{x})}_{\prod_J\cM_j}\) such that the following diagram commutes.

\begin{center}        
\begin{tikzpicture}[baseline=(current bounding box.center),
    transform shape, scale=.85]
\node[inner sep=0, outer sep=0]{
\begin{tikzcd}
	&& {\set{\vec{x} \mid r(\vec{t})(\vec{x})}_{\prod_J \cM_j}} & {\set{\vec{x} \mid r(\vec{t})(\vec{x})}_{\prod_F \cM}} & \terminal \\
	U & {(\prod_I M_i)_{\vec{x}}} & {(\prod_J M_j)_{\vec{x}}} & {(\prod_F M)_{\vec{x}}} & \Omega
	\arrow["\begin{array}{c} \\nu_J \end{array}", from=1-3, to=1-4]
	\arrow["{ \sem[\prod_J \cM_j]{\vec{x}.r(\vec{t})}}", tail, from=1-3, to=2-3]
	\arrow[from=1-4, to=1-5]
	\arrow["{ \sem[\prod_F \cM]{\vec{x}.r(\vec{t})}}", tail, from=1-4, to=2-4]
	\arrow["true", tail, from=1-5, to=2-5]
	\arrow["{\delta_J}", from=2-1, to=1-3]
	\arrow["{m }", curve={height=-45pt}, from=2-1, to=1-4] %
	\arrow["{{\alpha}}"', from=2-1, to=2-2]
	\arrow["{{(p_{I,J})_{\vec{x}}}}"', from=2-2, to=2-3]
	\arrow["{(\mu_I)_{\vec{x}}}"', shift right, curve={height=20pt}, from=2-2, to=2-4]
	\arrow["{{(\mu_J)_{\vec{x}}}}"', from=2-3, to=2-4]
	\arrow[from=2-4, to=2-5]
\end{tikzcd}
};
\end{tikzpicture}
\end{center}

The diagram shows that \(\prod_J \cM_j \models_{(p_{I,J})_{\vec{x}}\circ\alpha} \vec{x}.r(\vec{t})\). Since atomic formulae are interpreted componentwise in products, it follows that
\[
    \cM_j \models_{(p_{I,j})_{\vec{x}}\circ\alpha} \vec{x}.r(\vec{t})
\]
for every \(j\in J\). Thus, \(J \subseteq \SatisIndexSet{\vec{x}.r(\vec{t})}\). Since \(J\in F\) and \(F\) is a filter and, therefore, upward closed, we conclude that \(\SatisIndexSet{\vec{x}.r(\vec{t})}\in F\).

\item[\ref{item:atomic-products}.] Suppose that \(J = \SatisIndexSet{\vec{x}.r(\vec{t})}\) is in \(F\). By definition of \(\SatisIndexSet{\vec{x}.r(\vec{t})}\), for every \(j\in J\), \(\cM_j \models_{(p_{I,j})_{\vec{x}}\circ\alpha} \vec{x}.r(\vec{t})\). By the universal property of products, it follows that \(\prod_J\cM_j \models_{(p_{I,J})_{\vec{x}}\circ\alpha} \vec{x}.r(\vec{t})\). Let
\[
    m_J\from U \to \set{\vec{x}\mid r(\vec{t})(\vec{x})}_{\prod_J\cM_j}
\]
be a witness of this satisfaction. By Corollary~\ref{cor:atomic-commute-filtered-products}, the interpreting subobject \(\sem[\prod_F \cM]{\vec{x}.r(\vec{t})}\) of \(\vec{x}.r(\vec{t})\) in \(\prod_F\cM\) is the filtered product of the interpreting subobjects in the structures \(\cM_i\), with \(\nu\) the colimit. Henre, the diagram
\[\begin{tikzcd}
    {\set{\vec{x} \mid r(\vec{t})(\vec{x})}_{\prod_J \cM_j}} & {\set{\vec{x} \mid r(\vec{t})(\vec{x})}_{\prod_F \cM}} \\
    {(\prod_J M_j)_{\vec{x}}} & {(\prod_F M)_{\vec{x}}}
    \arrow["{\nu_J}", from=1-1, to=1-2]
    \arrow["{ \sem[\prod_J \cM_j]{\vec{x}.r(\vec{t})}}"', tail, from=1-1, to=2-1]
    \arrow["{ \sem[\prod_F \cM]{\vec{x}.r(\vec{t})}}", tail, from=1-2, to=2-2]
    \arrow["{{(\mu_J)_{\vec{x}}}}"', from=2-1, to=2-2]
\end{tikzcd}\]
commutes. Composing \(m_J\) with \(\nu_J\) yields a morphism 
\[
m = \nu_J \circ m_J \from U \to \set{\vec{x}\mid r(\vec{t})(\vec{x})}_{\prod_F\cM}.
\]
By commutativity of the previous diagram, \(\sem[\prod_F\cM]{\vec{x}.r(\vec{t})} \circ m = (\mu_I)_{\vec{x}}\circ\alpha\). Therefore \(\prod_F \cM \models_{(\mu_I)_{\vec{x}} \circ \alpha} \vec{x}.r(\vec{t})\).

\item[\ref{item:conjunction}.] Closure under finite conjunctions for \(\alpha\)-product-stable and \(\alpha\)-factor-stable formulae follows from Theorem~\ref{thm:fundamental theorem of KJ semantics} and closure properties of filters. More precisely, by Theorem~\ref{thm:fundamental theorem of KJ semantics}, we have the equality
\[
\SatisIndexSet{\vec{x}.(\varphi \land \psi)} = \SatisIndexSet{\vec{x}.\varphi} \cap \SatisIndexSet{\vec{x}.\psi}.
\]
Since \(F\) is a filter, it is both upward closed and closed under finite intersections, meaning that \(\SatisIndexSet{\vec{x}.(\varphi \land \psi)}\) is in \(F\) if and only if both \(\SatisIndexSet{\vec{x}.\varphi}\) and \(\SatisIndexSet{\vec{x}.\psi}\) are in \(F\). In particular, assume that both \(\psi\) and \(\varphi\) are \(\alpha\)-factor-stable. If \(\prod_F \cM \models_{(\mu_I)_{\vec{x}} \circ \alpha} \vec{x}.(\varphi \land \psi)\), then, by Theorem~\ref{thm:fundamental theorem of KJ semantics}, \(\prod_F \cM \models_{(\mu_I)_{\vec{x}} \circ \alpha} \vec{x}.\varphi\) and \(\prod_F \cM \models_{(\mu_I)_{\vec{x}} \circ \alpha} \vec{x}.\psi\). By stability under filtered factors, \(\SatisIndexSet{\vec{x}.\varphi}\) and \(\SatisIndexSet{\vec{x}.\psi}\) are in \(F\). We conclude that \(\varphi \land \psi\) is \(\alpha\)-factor-stable. The same proof, decomposing in the reserve direction, gives the closure under finite conjunctions for the class of \(\alpha\)-product-stable formulae. It can also be seen as a special case of the next item.

\item[\ref{item:infinite-conjunction}.] Consider a family \((\varphi_k)_{k \in K}\) of \(\alpha\)-product-stable formulae. Suppose that \(\SatisIndexSet{\vec{x}.\bigwedge_K \varphi_k} \in F\).
By Theorem~\ref{thm:fundamental theorem of infinitary KJ semantics}, for every \(k \in K\), \(\SatisIndexSet{\vec{x}.\bigwedge_K \varphi_k} \subseteq \SatisIndexSet{\vec{x}.\varphi_k}\). Since filters are upward closed, it follows that \(\SatisIndexSet{\vec{x}.\varphi_k} \in F\) for every \(k\).
By stability under filtered products, it follows that, for every \(k \in K\), \(\prod_F \cM \models_{(\mu_I)_{\vec{x}} \circ \alpha} \vec{x}.\varphi_k\). By Theorem~\ref{thm:fundamental theorem of infinitary KJ semantics}, we conclude that \(\prod_F \cM \models_{(\mu_I)_{\vec{x}} \circ \alpha} \vec{x}.\bigwedge_K \varphi_k\) which gives the closure under arbitrary conjunctions for the class of \(\alpha\)-product-stable formulae.

By Theorem~\ref{thm:fundamental theorem of infinitary KJ semantics}, for every \(k \in K\), \(\SatisIndexSet{\vec{x}.\bigwedge_K \varphi_k} \subseteq \SatisIndexSet{\vec{x}.\varphi_k}\). Since filters are upward closed, it follows that \(\SatisIndexSet{\vec{x}.\varphi_k} \in F\) for every \(k\).

\item[\ref{item:disjunction-filters}.] Let \(\varphi\) and \(\psi\) be two factor-stable formulae and suppose that
\[
\prod_F \cM \models_{(\mu_I)_{\vec{x}} \circ \alpha} \vec{x}.(\varphi \lor \psi).
\]
By \cref{thm:fundamental theorem of KJ semantics}, there are two morphisms $p\from V \to U$ and $q\from W \to U$ such that
\begin{itemize}
    \item $p+q$ is an epimorphism,
    \item $\prod_F \cM \models_{(\mu_I)_{\vec{x}} \circ \alpha \circ p} \vec{x}.\varphi$, and
    \item $\prod_F \cM \models_{(\mu_I)_{\vec{x}} \circ \alpha \circ q} \vec{x}.\psi$.
\end{itemize}
By stability under filtered product at \(\alpha \circ p\), resp. \(\alpha \circ q\), it follows that \(\SatisIndexSet[\alpha \circ p]{\vec{x}.\varphi}\) and \(\SatisIndexSet[\alpha \circ q]{\vec{x}.\psi}\) are in \(F\). Because \(F\) is a filter,
\[
J = \SatisIndexSet[\alpha \circ p]{\vec{x}.\varphi} \cap \SatisIndexSet[\alpha \circ q]{\vec{x}.\psi} \in F.
\]
Now, for \(j \in J\), \(\cM_j \models_{(p_{I,j})_{\vec{x}} \circ \alpha \circ p} \vec{x}.\varphi\) and \(\cM_j \models_{(p_{I,j})_{\vec{x}} \circ \alpha \circ q} \vec{x}.\psi\) and thus, by \cref{thm:fundamental theorem of KJ semantics}, \(\cM_j \models_{(p_{I,j})_{\vec{x}} \circ \alpha} \vec{x}.(\varphi \lor \psi)\). Since \(F\) is upward-closed as a filter, \(\SatisIndexSet[\alpha]{\vec{x}.(\varphi \lor \psi)}\) is also in \(F\).

\item[\ref{item:implication}.] Suppose that \(\varphi\) is factor-stable and \(\psi\) is product-stable. We show that \(\varphi \Rightarrow \psi\) is product-stable by contraposition and assume \(\prod_F \cM \not\models_{(p_{I,i})_{\vec{x}} \circ \alpha} \vec{x}.\varphi \Rightarrow \psi\). By Theorem~\ref{thm:fundamental theorem of KJ semantics}, there exists $p\from V \to U$ such that 
\begin{itemize}
    \item $\prod_F \cM \models_{(\mu_I)_{\vec{x}} \circ \alpha \circ p} \vec{x}.\varphi$
    \item $\prod_F \cM \not\models_{(\mu_I)_{\vec{x}} \circ \alpha \circ p} \vec{x}.\psi$
\end{itemize}
Since \(\varphi\) is factor-stable, it follows that \(\SatisIndexSet[\alpha \circ p]{\vec{x}.\varphi}\) is in \(F\). Similarly, since \(\psi\) is product-stable, we have that \(\SatisIndexSet[\alpha \circ p]{\vec{x}.\psi}\) is not in \(F\). Since \(F\) is an ultrafilter, \(I \setminus \SatisIndexSet[\alpha \circ p]{\vec{x}.\psi}\) is in \(F\). Since \(F\) is closed by intersection,
\[
\SatisIndexSet[\alpha \circ p]{\vec{x}.\varphi} \cap (I \setminus \SatisIndexSet[\alpha \circ p]{\vec{x}.\psi}) \in F.
\]
By Theorem~\ref{thm:fundamental theorem of KJ semantics}, it follows that
\[
\SatisIndexSet[\alpha \circ p]{\vec{x}.\varphi} \cap (I \setminus \SatisIndexSet[\alpha \circ p]{\vec{x}.\psi}) \subseteq \set{i \in I \mid \cM_i \not\models_{(p_{I,i})_{\vec{x}} \circ \alpha} \vec{x}.\varphi \Rightarrow \psi}.
\]
Thus, \(\set{i \in I \mid \cM_i \not\models_{(p_{I,i})_{\vec{x}} \circ \alpha} \vec{x}.\varphi \Rightarrow \psi}\) is in \(F\), and then
\[
I \setminus \set{i \in I \mid \cM_i \not\models_{(p_{I,i})_{\vec{x}} \circ \alpha} \vec{x}.\varphi \Rightarrow \psi} = \SatisIndexSet{\vec{x}.(\varphi \Rightarrow \psi)}
\] is not in \(F\). We conclude, by contraposition, that \(\varphi \Rightarrow \psi\) is product-stable.

\item[\ref{item:negation-factor-to-product}.] Suppose that \(\varphi\) is factor-stable. We prove that \(\neg \varphi\) is product-stable by contraposition.
Thus, we suppose that
\[
    \prod_F \not\models_{(\mu_I)_{\vec{x}} \circ \alpha} \vec{x}.\neg \varphi.
\]
By Theorem~\ref{thm:fundamental theorem of KJ semantics}, there exists a morphism $p\from V \to U$ with $V \not\simeq \emptyset$ such that $\prod_F \cM \models_{(\mu_I)_{\vec{x}} \circ \alpha \circ p} \vec{x}.\varphi$. Because \(\varphi\) is factor-stable, the set \(\SatisIndexSet[\alpha \circ p]{\vec{x}.\varphi}\) is in \(F\), and then, by Theorem~\ref{thm:fundamental theorem of KJ semantics} and upward closure of \(F\) as a filter, the set \(\set{i \in I \mid \cM_i \not\models_{(p_{I,i})_{\vec{x}} \circ \alpha} \vec{x}.\neg \varphi}\) is also in \(F\). As $F$ is an ultrafilter, we then have that
\[
    \SatisIndexSet{\vec{x}.\neg \varphi} \notin F.
\]
Therefore, \(\neg \varphi\) is product-stable.

\item[\ref{item:existential-quantification-factors}.] Suppose that \(\varphi\) is factor-stable at every generalized element with a presentable domain. To show that that \(\exists y. \varphi\) is factor-stable (with \(y\from s\)), let us assume that \(\prod_F \cM \models_{(\mu_I)_{\vec{x}} \circ \alpha} \vec{x}.(\exists y. \varphi)\).
By Proposition~\ref{prop:existential and representable object}, there exists an epimorphism \(p\from V \to U\) with presentable domain \(V\) and a generalized element \(\beta\from V \to (\prod_F M)_s\) such that \(\prod_F \cM \models_{((\mu_I)_{\vec{x}} \circ \alpha \circ p,\beta)} \vec{x}y.\varphi\).
From Proposition~\ref{prop:lifting-filtered-products}, $\beta$ factorizes through $(\mu_I)_s$, i.e., there exists $\delta\from V \to (\prod_I M_i)_s$ such that $\beta = (\mu_I)_s \circ \delta$.
Then,
\[
    \prod_F \cM \models_{(\mu_I)_{\vec{x}s} \circ (\alpha \circ p, \delta)} \vec{x}y.\varphi.
\]
Since \(\varphi\) is in \(\cF_{\mathrm{pr}}\), \(\SatisIndexSet[\alpha \circ p, \delta]{\varphi} \in F\).
Let \(j\) in \(\SatisIndexSet[\alpha \circ p, \delta]{\varphi}\).
Then,
\[
    \cM_j \models_ {(p_{I,j})_{\vec{x}s} \circ (\alpha \circ p,\delta)} \vec{x}y.\varphi,
\]
where \(p\from V \to U\) is an epimorphism with presentable domain \(V\), and \((p_{I,j})_s \circ \delta \from V \to (M_j)_s\) is a generalized element.
By Proposition~\ref{prop:existential and representable object}, it follows that \(\cM_j \models_{(p_{I,j})_{\vec{x}} \circ \alpha} \vec{x}.(\exists y. \varphi)\).
Thus, it holds that
\[
    \SatisIndexSet[\alpha \circ p, \delta]{\varphi} \subseteq \SatisIndexSet{\exists y. \varphi}.
\]
Since \(F\) is a filter, we can conclude that \(\SatisIndexSet{\exists y. \varphi} \in F\). 

\item[\ref{item:existential-quantification-products}.]
Suppose that \(\varphi\) is product-stable at every global element (i.e., whenever \(U = \terminal\)), and that \(\terminal\) is projective. To show that that \(\exists y. \varphi\) is product-stable (with \(y\from s\)) at every global element, let us assume that \(J = \SatisIndexSet{\vec{x}.(\exists y.\varphi)}\) is in \(F\).
For every \(j \in J\), Theorem~\ref{thm:fundamental theorem of KJ semantics} gives a generalized element \(\beta_j\from V_j \to (M_j)_s\), with \(V_j\) internally inhabited, such that
\[
    \cM_j \models_{((p_{I,j})_{\vec{x}} \circ \alpha \circ !_{V_j},\beta_j)} \vec{x}y.\varphi.
\]
Since \(\terminal\) is projective and \(V_j\) is internally inhabited, \(!_{V_j}\) admits a global section \(s_j\from \terminal \to V_j\), with \(!_{V_j} \circ s_j = \id_\terminal\).
By monotonicity (see Proposition~\ref{prop:monotonicity}), it follows that, for every \(j \in J\),
\begin{equation}
    \label{eq:satisfaction:at:j:existential}
    \cM_j \models_{((p_{I,j})_{\vec{x}} \circ \alpha,\, \beta_j \circ s_j)} \vec{x}y.\varphi.
\end{equation}
Define \(\tilde{\beta}_j = \beta_j \circ s_j\).
Besides, \((M_i)_s\) is internally inhabited for every \(i \in I \setminus J\) and thus admits a global section \(\tilde{\beta}_i\from \terminal \to (M_i)_s\). 
By the universal property of product, there is a unique morphism \(\beta\from \terminal \to (\prod_I M_i)_s\) such that for every \(j \in J\), \((p_{I,j})_s \circ \beta = \tilde{\beta}_j\).
Now, we have \(\beta_j \circ s_j = \tilde{\beta}_j = (p_{I,j})_s \circ \beta\). Since products commute componentwise,
\[
    ((p_{I,j})_{\vec{x}} \circ \alpha,\, \beta_j \circ s_j) = (p_{I,j})_{\vec{x}s} \circ (\alpha, \beta).
\]
Thus, Equation~(\ref{eq:satisfaction:at:j:existential}) entails
\(
    J \subseteq \SatisIndexSet[(\alpha, \beta)]{\vec{x}y.\varphi}
\).
Since \(J\in F\) and \(F\) is a filter and, therefore, upward closed, we deduce that \(\SatisIndexSet[(\alpha, \beta)]{\vec{x}y.\varphi}\in F\). As \(\varphi\) is product-stable, it follows that
\[
    \prod_F \cM \models_{(\mu_I)_{\vec{x}s} \circ (\alpha, \beta)} \vec{x}y.\varphi.
\]
Theorem~\ref{thm:fundamental theorem of KJ semantics} implies that \(\prod_F \cM \models_{(\mu_I)_{\vec{x}} \circ \alpha} \vec{x}.(\exists y. \varphi)\) and we conclude that \(\exists y. \varphi\) is product-stable at every global element.

\item[\ref{item:universal-quantification-factor}.]
Suppose that \(\varphi\) is factor-stable. We reason by contraposition to show that \(\forall y.\varphi\) (with \(y\from s\)) is also factor-stable and suppose that \(\SatisIndexSet{\vec{x}.\forall y.\varphi} \not \in F\). Let
\[
J = \set{i \in I \mid \cM_i \not\models_{(p_{I,i})_{\vec{x}} \circ \alpha} \vec{x}.(\forall y. \varphi)}.
\]
Since \(F\) is an ultrafilter, it follows that \(J\) is in \(F\).
For every for all \(j \in J\), Theorem~\ref{thm:fundamental theorem of KJ semantics} implies that there exists a morphism $p_j\from V_j \to U$ and a generalized element $\beta_j\from V_j \to (M_j)_s$ such that 
\(\cM_j \not\models_{((p_{I,j})_{\vec{x}} \circ \alpha \circ p_j,\beta_j)} \vec{x}y.\varphi\).
Since the family \(((M_i)_s)_{i \in I}\) is internally inhabited and \(\terminal\) is projective, Corollary~\ref{cor:lifting-from-projective-terminal} ensures that the family is lifting along products. From Definition~\ref{def:lifting-along-products}, $\beta_j$ factors through the canonical projection \((p_{I,j})_s\), i.e., there exists a morphism $\delta_j\from V_j \to (\prod_I \cM_i)_s$ such that $\beta_j = (p_{I,j})_s \circ \delta_j$, and we obtain
\[
    \cM_j \not\models_{(p_{I,j})_{\vec{x}s} \circ (\alpha \circ p_j,\delta_j)} \vec{x}y.\varphi.
\]
Let \(V = \coprod_{j\in J} V_j\) with canonical injections $i_j\from V_j \to V$. Then, by the universal property of colimit, there is a unique morphism $q\from V \to U$ such that $p_j = q \circ i_j$. Likewise, there is a unique morphism $\delta\from V \to (\prod_I M_i)_s$ such that $\delta_j = \delta \circ i_j$. Hence, we have that, for all \(j \in J\),
\[
    \cM_j \not\models_{(p_{I,j})_{\vec{x}s} \circ (\alpha \circ q \circ i_j,\delta \circ i_j)} \vec{x}y.\varphi,
\]
and then by contraposition of monotonicity (Proposition~\ref{prop:monotonicity}), for all \(j \in J\),
\[
    \cM_j \not\models_{(p_{I,j})_{\vec{x}s} \circ (\alpha \circ q,\delta)} \vec{x}y.\varphi.
\]
Hence, \(J \subseteq (I \setminus \SatisIndexSet[\alpha \circ q,\delta]{\varphi})\).
Since \(J \in F\) and \(F\) is an ultrafilter, \(\SatisIndexSet[\alpha \circ q,\delta]{\varphi}\) is not in \(F\). Because \(\varphi\) is factor-stable, we deduce that
\[
    \prod_F \cM \not\models_{(\mu_I)_{\vec{x}s} \circ (\alpha \circ q,\delta)} \vec{x}y.\varphi.
\]
By the Kripke--Joyal clause for universal quantification (Theorem~\ref{thm:fundamental theorem of KJ semantics}), we conclude that $\prod_F \cM \not\models_{(\mu_I)_{\vec{x}} \circ \alpha} \vec{x}.\forall y.\varphi$.

\item[\ref{item:universal-quantification-product}.]
Suppose that \(\varphi\) is product-stable. We reason by contraposition to show that \(\forall y.\varphi\) (with \(y\from s\)) is also product-stable and suppose that
\[
    \prod_F \cM \not\models_{(\mu_I)_{\vec{x}} \circ \alpha} \vec{x}.(\forall y.\varphi).
\]
By Theorem~\ref{thm:fundamental theorem of KJ semantics}, there exists a morphism $p\from V \to U$ and a generalized element $\beta\from V \to (\prod_F M)_s$ such that $\prod_F \cM \not\models_{((\mu_I)_{\vec{x}} \circ \alpha \circ p,\beta)} \vec{x}y.\varphi$.
Since the family \(((M_i)_s)_{i \in I}\) is internally inhabited and \(\terminal\) is projective, Corollary~\ref{cor:lifting-from-projective-terminal} ensures that the family is lifting along products.
From Proposition~\ref{prop:lifting-filtered-products}, $\beta$ factorizes through $(\mu_I)_s$, i.e., there exists a morphism $\delta\from V \to (\prod_I \cM_i)_s$ such that $\beta = (\mu_I)_s \circ \delta$.
Thus, we obtain
\[
    \prod_F \cM \not\models_{(\mu_I)_{\vec{x}s} \circ (\alpha \circ p,\delta)} \vec{x}y.\varphi.
\]
Because \(\varphi\) is product-stable, it follows that 
\(
    \SatisIndexSet[\alpha \circ p,\delta]{\vec{x}y.\varphi} \notin F
\).
Since $F$ is an ultrafilter, the set \(J = \set{i \in I \mid \cM_i \not\models_{(p_{I,i})_{\vec{x}y} \circ (\alpha \circ p,\delta)} \vec{x}y.\varphi}\) is in \(F\).
Thus, for any \(j \in J\),
\[
    \cM_j \not\models_{(p_{I,j})_{\vec{x}s} \circ (\alpha \circ p,\delta)} \vec{x}y.\varphi,
\]
with $p$ a morphism $V \to U$ and \((p_{I,j})_s \circ \delta\) a generalized element \(V \to (M_j)_s\).
By Theorem~\ref{thm:fundamental theorem of KJ semantics}, we obtain that \(\cM_j \not\models_{(p_{I,j})_{\vec{x}} \circ \alpha} \vec{x}.(\forall y.\varphi)\).
In particular,
\[
J \subseteq \set{i \in I \mid \cM_i \not\models_{(p_{I,i})_{\vec{x}} \circ \alpha} \vec{x}.(\forall y.\varphi)}.
\]
As \(F\) is an ultrafilter, we deduce that
\(
    \set{i \in I \mid \cM_i \not\models_{(p_{I,i})_{\vec{x}} \circ \alpha} \vec{x}.(\forall y.\varphi)} \in F
\)
and conclude that \(\SatisIndexSet{\vec{x}.(\forall y.\varphi)} \notin F\).

\item[\ref{item:disjunction-products}.]

Assume that \(\terminal\) is projective, and \(\cC\) is two-valued.
Let \(\varphi\) and \(\psi\) be two formulae, product-stable at the global element \(\alpha\from \terminal \to (\prod_I M_i)_{\vec{x}}\). Suppose that \(J = \SatisIndexSet[\alpha]{\vec{x}.(\varphi \lor \psi)}\) belongs to \(F\). Then,
\(
\cM_j \models_{(p_{I,j})_{\vec{x}} \circ \alpha} \vec{x}.(\varphi \lor \psi).
\)
for every \(j\in J\). Theorem~\ref{thm:fundamental theorem of KJ semantics} gives, for every \(j \in J\), objects \(V_j\) and \(W_j\) such that:
\begin{itemize}
    \item \(!_{V_j + W_j}\from V_j + W_j \to \terminal\) is an epimorphism,
	\item $\cM_j \models_{(p_{I,j})_{\vec{x}} \circ \alpha \circ !_{V_j}} \vec{x}.\varphi$
	\item $\cM_j \models_{(p_{I,j})_{\vec{x}} \circ \alpha \circ !_{W_j}} \vec{x}.\psi$
\end{itemize}

Now, \(V_j+W_j\) being internally inhabited ensures that \(V_j+W_j\not\simeq\initial\), meaning that at least one of \(V_j\) and \(W_j\) is non-initial (because \(\initial\) is strict in a topos).

Assume that \(V_j \not\simeq \initial\) for some \(j\in J\). Since \(\cC\) is two-valued, it follow that \(V_j\) is internally inhabited and \(!_{V_j}\) is an epimorphism. Because \(\terminal\) is projective, \(!_{V_j}\) admits a global section \(s_j\from \terminal \to V_j\). By monotonicity (Proposition~\ref{prop:monotonicity}), we obtain that 
\[
    \cM_j \models_{(p_{I,j})_{\vec{x}} \circ \alpha} \vec{x}.\varphi.
\]
Similarly, if \(W_j \not\simeq \initial\), it follows that \(\cM_j \models_{(p_{I,j})_{\vec{x}} \circ \alpha} \vec{x}.\psi.\)

Since at least one of \(V_j\) and \(W_j\) is non-initial, for every \(j\in J\), at least one of \(\cM_j \models_{(p_{I,j})_{\vec{x}} \circ \alpha} \vec{x}.\varphi\) and \(\cM_j \models_{(p_{I,j})_{\vec{x}} \circ \alpha} \vec{x}.\psi\) holds. Define
\begin{itemize}
    \item \(J_\varphi = \set{j\in J \mid \cM_j \models_{(p_{I,j})_{\vec{x}} \circ \alpha} \vec{x}.\varphi}\) and 
    \item \(J_\psi = \set{j\in J \mid \cM_j \models_{(p_{I,j})_{\vec{x}} \circ \alpha} \vec{x}.\psi}\).
\end{itemize}
Then
\[
    J = J_\varphi \cup J_\psi.
\]
Since \(J\in F\) and \(F\) is an ultrafilter, it follows that \(J_\varphi \in F\) or \(J_\psi \in F\). Without loss of generality, assume that \(J_\varphi \in F\). Then \(\SatisIndexSet{\vec{x}.\varphi} \in F\). Since \(\varphi\) is product-stable at \(\alpha\), it follows that \(\prod_F\cM \models_{(\mu_I)_{\vec{x}}\circ\alpha} \vec{x}.\varphi\). The reserve direction for the disjunction in Theorem~\ref{thm:fundamental theorem of KJ semantics} (taking \(\initial\) for the satisfaction of \(\psi\)) gives 
\[
    \prod_F\cM \models_{(\mu_I)_{\vec{x}}\circ\alpha} \vec{x}.(\varphi\lor\psi).
\]
Thereafter, \(\varphi\lor\psi\) is product-stable at the global element \(\alpha\).

\item[\ref{item:negation-factor}.] 
    
Suppose \(\terminal\) is projective, \(F\) is an ultrafilter, and \(\cC\) is two-valued.
Let \(\varphi\) be a formula factor-stable at a global element \(\alpha\from \terminal \to (\prod_I M_i)_{\vec{x}}\).
We prove that \(\neg \varphi\) is factor-stable at \(\alpha\) by contraposition, and suppose that \(\SatisIndexSet{\vec{x}.\neg \varphi} \notin F\).

Since \(F\) is an ultrafilter, the set \(K = I \setminus \SatisIndexSet{\vec{x}.\neg \varphi} \) belongs in \(F\). In particular, for every \(k \in K\), \(\cM_k \not\models_{(p_{I,k})_{\vec{x}} \circ \alpha} \vec{x}.\neg \varphi\).
By \cref{thm:fundamental theorem of KJ semantics}, for every \(k \in K\), there is \(V_k \neq \initial\) such that
\begin{equation}
    \label{eq:satisfaction:at:k:negation}
    \cM_k \models_{(p_{I,k})_{\vec{x}} \circ \alpha \circ !_{V_k}} \vec{x}.\varphi.
\end{equation}

Since \(\cC\) is two-valued, every subterminal object is either \(\initial\) or \(\terminal\). In particular, the image \(\im(!_{V_k})\) is either \(\initial\) or \(\terminal\). However, as \(V_k \neq \initial\), it follows that \(\im(!_{V_k}) = \terminal\). Thus \(V_k\) is internally inhabited for every \(k \in K\). By Proposition~\ref{prop: products preserve inhabitedness}, $\prod_K V_k$ is also internally inhabited.

Now, consider the family \(V = (V_k)_{k \in K}\) and let \(F_{\mid K} = \set{J \cap K \mid J \in F}\) denote the filter induced by \(F\) on \(K\). It follows from Proposition~\ref{prop:filtered product is inhabited} that \(\prod_{F_{\mid K}}\!\!V\) is internally inhabited. In particular, \(\prod_{F_{\mid K}}\!\!V \neq \initial\).

For every \(k\in K\), the universal properties of products and filtered products give
\[
    !_{\prod_{F_{\mid K}}\!\!V} \circ \mu^V_K {}={} !_{V_k} \circ \pi^V_{K,k}.
\]
where \(\pi^V_{K,k} \from \prod_K V_k \to V_k\) is the canonical projection and \(\mu^V_K \from \prod_K V_k \to \prod_{F_{\mid K}}\!\!V\) is the colimit morphisms.
Applying monotonicity (Proposition~\ref{prop:monotonicity}) to Equation~(\ref{eq:satisfaction:at:k:negation}) ensures that, for every \(k \in K\),
\[
\cM_k \models_{(p_{I,k})_{\vec{x}} \circ \alpha \circ !_{V_k} \circ \pi^V_{K,k}} \vec{x}.\varphi,
\]
and then
\[
\cM_k \models_{(p_{I,k})_{\vec{x}} \circ \alpha \circ !_{\prod_{F_{\mid K}}\!\!V} \circ \mu^V_K} \vec{x}.\varphi.
\]

Since $\prod_K V_k$ is internally inhabited and \(\terminal\) is projective, there exists a global section $s\from 1 \to \prod_K V_k$, and then $!_{\prod_{F_{\mid K}}\!\!V} \circ \mu^V_K \circ s = \id_\terminal$. Hence, we have that 
\[
\cM_k \models_{(p_{I,k})_{\vec{x}} \circ \alpha} \vec{x}.\varphi.
\]

Thus, \(K \subseteq \SatisIndexSet{\vec{x}.\varphi}\). Since \(K\in F\) and \(F\) is a filter and, therefore, upward closed, we deduce that \( \SatisIndexSet{\vec{x}.\varphi}\in F\).
Because \(\varphi\) is factor-stable at \(\alpha\), it follows that
\[
\prod_F \cM \models_{(\mu_I)_{\vec{x}} \circ \alpha} \vec{x}.\varphi.
\]
In particular, it follows that
\[
\prod_F \cM \not\models_{(\mu_I)_{\vec{x}} \circ \alpha} \vec{x}.\neg\varphi,
\]
and we can conclude that \(\neg \varphi\) is factor-stable at the global element \(\alpha\).

\end{enumerate}
\end{proof}

Combining the previous closure properties yields the following generalized form of \Los's theorem.

\begin{Corollary}[Generalized \Los's theorem]
\label{cor:Los theorem}
Let \(\cC\) be a locally presentable topos, $(\cM_i)_{i \in I}$ be a family of internally inhabited $\Sigma$-structures, and $F$ be an ultrafilter on $I$.

\begin{enumerate}
\item {\em Cartesian logic.} If $\varphi$ is a Cartesian formula (with suitable context \(\vec{x}\)), i.e. a formula built from atomic formulae using finite conjonctions, then, for every generalized element $\alpha\from U \to (\prod_I M_i)_{\vec{x}}$
\[
    \SatisIndexSet{\vec{x}.\varphi} \in F
    \; \text{iff} \;
    \prod_F \cM \models_{(\mu_I)_{\vec{x}} \circ \alpha} \vec{x}.\varphi.
\]

\item {\em Horn logic} If $\varphi$ and $\psi$ are two Cartesian formulae (with suitable context \(\vec{x}\)), then, for every generalized element $\alpha\from U \to (\prod_I M_i)_{\vec{x}}$
\[
    \SatisIndexSet{\vec{x}.(\varphi \Rightarrow \psi)} \in F
    \; \text{implies} \;
    \prod_F \cM \models_{(\mu_I)_{\vec{x}} \circ \alpha} \vec{x}.\varphi \Rightarrow \psi.
\]
\end{enumerate}
Now, suppose, in addition, that $\terminal$ is projective and that \(\cC\) is two-valued.
\begin{enumerate}[resume]
\item {\em Coherent logic} If $\varphi$ and $\psi$ are two coherent sentences, 
i.e., sentences built from atomic formulae using finite conjunctions, finite disjunctions, and existential quantification,\footnote{
Or, more generally,
built from atomic formulae using arbitrary conjunctions, finite disjunctions, existential and universal quantification.
}
then, for all every element $\alpha\from \terminal \to (\prod_I M_i)_{\vec{x}}$
\[
    \SatisIndexSet{[].(\varphi \Rightarrow \psi)} \in F
    \; \text{implies} \;
    \prod_F \cM \models_{(\mu_I)_{[]} \circ \alpha} [].\varphi \Rightarrow \psi.
\]
    
\item\label{case:weak-FOL:cor:Los theorem} {\em Weak first-order logic} If $\varphi$ is a formula (with suitable context \(\vec{x}\)) built from atomic formulae
using conjunctions, negations, and universal quantification,
then, for every global element $\alpha\from \terminal \to (\prod_I M_i)_{\vec{x}}$
\[
    \SatisIndexSet{\vec{x}.\varphi} \in F
    \; \mbox{iff} \;
    \prod_F \cM \models_{(\mu_I)_{\vec{x}} \circ \alpha} \vec{x}.\varphi.
\]
\end{enumerate}
\end{Corollary}

\begin{proof}
The first two cases follow immediately from the closure properties of $\cF_{\mathrm{pr}}$, $\cP$, and $\cF$ established in Theorem~\ref{thm:Los:kjsem}. For the coherent case, sentences have empty context, so there is no choice of generalized element: their satisfaction is evaluated at the unique global element of the terminal object. The result follows by combining the closure properties for atomic formulae, finite conjunctions, finite disjunctions, and existential quantification. The weak first-order case is obtained similarly from the closure properties for conjunction, negation, and universal quantification.
\end{proof}

If \(\cC\) is Boolean, then \(\set{\neg, \land, \forall}\) is a complete set of connectives for first-order logic. 
In particular, the previous weak first-order case (property~\ref{case:weak-FOL:cor:Los theorem}) yields the classical \Los{} equivalence for sentences and, more generally, for formulae evaluated at global elements:
\[
    \SatisIndexSet{\vec{x}.\varphi} \in F
    \; \mbox{iff} \;
    \prod_F \cM \models_{(\mu_I)_{\vec{x}} \circ \alpha} \vec{x}.\varphi.
\]
In particular, we recover the classical set-theoretic version of \Los's theorem.
Indeed \(\Set\) is a locally finitely presentable elementary two-valued Boolean topos where $\terminal$ is projective, and an \(I\)-sequence \((a_i)_{i \in I}\) in \(\Set\) is precisely a global element \(\alpha \from \terminal \to (\prod_I M_i)_{\vec{x}}\). Thus, applying Corollary~\ref{cor:Los theorem} to such global elements \(\alpha\) yields \cref{thm:Los:set-theory}.

\subsection{Counterexamples and a Conjecture on \Los's Property}
\label{subsec:los:hypotheses}

The hypotheses in Theorem~\ref{thm:Los:kjsem} reflect the different obstacles encountered in its proof,
In particular, two-valuedness and projectivity of the terminal object allows turning certain local witnesses into global one.
These assumptions should, therefore, be considered as sufficient conditions for our proof, rather than as necessary for a topos to satisfy \Los's theorem.
More precisely, consider the following \emph{\Los's property}:

\begin{Definition}[\Los's property]
A locally presentable topos \(\cC\) has {\bf \Los's property} if, for every signature \(\Sigma\), every family of inhabited \(\Sigma\)-structures \((\cM_i)_{i \in I}\) interpreted in \(\cC\), every ultrafilter \(F\) on \(I\), and every \(\Sigma\)-formula \(\vec{x}.\varphi\),
\[
\prod_F \cM \models_{(\mu_I)_{\vec{x}} \circ \alpha} \vec{x}.\varphi
\; \mbox{iff} \;
\SatisIndexSet{\vec{x}.\varphi} \in F
\]
for every global element \(\alpha\from \terminal \to (\prod_I M_i)_{\vec{x}}\).
\end{Definition}

We first show that two-valuedness and projectivity of the terminal object are not necessary for a topos to have \Los's property through two counterexamples: topos \(\Set \times \Set\) and \(G\text{-}\Set\).

\begin{Example}
\label{ex:setxset}
Consider the Boolean topos \(\Set \times \Set\). It is not two-valued, since \(\Omega \simeq \set{(0,0),(0,1),(1,0),(1,1)}\), but its terminal object is projective and it has \Los's property. Indeed, a \(\Sigma\)-structure \(\cM\) in \(\Set\times\Set\) is a pair \((\cM_1,\cM_2)\) of \(\Sigma\)-structures in \(\Set\). Global elements and filtered products are computed componentwise, and satisfaction of formulae is, likewise, componentwise:
\[
\cM \models_\alpha\vec{x}.\varphi
\; \mbox{iff} \; 
\cM_1\models_{\alpha_1}\vec{x}.\varphi
\; \mbox{and} \; 
\cM_2 \models_{\alpha_2}\vec{x}.\varphi,
\]
where \(\alpha_i\) is the corresponding component of \(\alpha\). Thus, the classical version of \Los's theorem applied to each component gives \Los's property in \(\Set\times\Set\).
\end{Example}

\begin{Example}
\label{ex:gset}
Consider the Boolean topos \(G\text{-}\Set\) of left actions of a non-trivial group \(G\) on sets. Its terminal object is the one-element set equipped with the trivial action. The canonical epimorphism map \(\pi \from G \to \terminal\), where \(G\) acts on itself by left translation, has no section. Indeed a section \(s \from \terminal \to G\) would give a fixed point of the action of \(G\) on itself, which is does not exist unless \(G = \set{e}\). In particular, \(\terminal\) is not projective in \(G\text{-}\Set\). Still, the forgetful functor \(G\text{-}\Set\to\Set\) creates limits and colimits. In particular, filtered products and filtered colimits are computed on the underlying sets, and satisfaction of formulas is reflected by the underlying set semantics. Here again, the classical \Los's theorem in \(\Set\) applies to the underlying structures and yields \Los's property in \(G\text{-}\Set\).
\end{Example}

These examples show that our assumptions provide an internal mechanism for obtaining the closure properties used to derive \Los's theorem. They also suggest an alternative, external perspective. Instead of building global elements from local witnesses, one could try to observe the internal semantics through suitable logical points \(p\from\cC\to\Set\), namely geometric morphisms whose inverse-image functors \(p^*\) additionally preserve the subobject classifier and exponentials.
Such logical points preserve and reflect the relevant logical semantics: for every \(\Sigma\)-structure \(\cM\) interpreted in \(\cC\), generalized element \(\alpha\from U \to M_{\vec{x}}\), and formula \(\vec{x}.\varphi\),
\begin{equation}
\label{eq:preservation through logical points}
\cM \models_\alpha \vec{x}.\varphi \; \mbox{iff} \; p^*(\cM) \models_{p^*(\alpha)} \vec{x}.\varphi.
\end{equation}
If the logical points are compatible with the ultraproduct construction, then the classical \Los's theorem in \(\Set\) can be transported back to \(\cC\). This suggests the following conjecture:

\begin{Conjecture}
\label{conjecture}
\Los's property for full FOL holds in locally presentable topos $\cC$ if, and only if $\cC$ admits a conservative family of logical points commuting with filtered colimits.
\end{Conjecture}

The sufficient direction is a direct consequence of the equivalence~(\ref{eq:preservation through logical points}). The necessary direction remains to be investigated. Establishing such an equivalence would provide an external counterpart to our internal proof. Classical representation results such as Deligne's theorem~\cite[Proposition 9.0, exposé VI, p.336]{verdier_theorie_1972} and Barr's theorem~\cite{barr_toposes_1974} suggest that such an external perspective may be natural, although whether these results can be used to characterize \Los's property remains open.

\subsection{Compactness Theorem}
\label{subsec:los:compactness}

A {\bf theory} \(T\) is a set of sentences.
We say that \(T\) is {\bf consistent with inhabited sorts} in \(\cC\) if it admits an internally inhabited model \(\cM\) in \(\cC\).

By the Kripke--Joyal semantics of existential quantification (see Theorem~\ref{thm:fundamental theorem of KJ semantics}), a structure \(\cM\) is internally inhabited if and only if
\[
    \cM \models \exists x\from s.\top
\]
for every sort \(s\) of \(\Sigma\). This strengthening the usual notion of consistency, i.e., the existence of a model, is necessary for our compactness argument as ordinary consistency does not guarantee that a model is internally inhabited in a topos. For example, consider
the theory
\[
    T = \set{\forall x\from s.\bot}
\]
in a signature containing a sort \(s\). This theory has a model in \(\cC\), by interpreting \(s\) as the initial object \(\initial\) of \(\cC\). However, an internally inhabited model of \(T\) would satisfy both \(\cM \models \exists x\from s.\top\), and \(\cM \models \forall x\from s.\bot\), and, therefore, \(\cM\models\bot\). In particular, ordinary consistency is too weak to apply Corollary~\ref{cor:Los theorem} and obtain the usual compactness result out of \Los's theorem.

We consider the following classes of logic:
\begin{itemize}

    \item {\bf Coherent logic}: formulae are built from atomic formulae using finite conjunction, finite disjunction, and existential  quantification. Coherent sequents are implications \(\varphi \Rightarrow \psi\) between such formulae.

    \item {\bf Positive logic}: formulae are built from atomic formulae using finite conjunction, finite disjunction, and universal and existential quantification. We similarly consider implications between such formulae.
    
    \item {\bf First-order logic (FOL)}: formulae are built from \(\bot\), \(\top\), and atomic formulae using finite conjunction and disjunction, implication, negation,and universal and existential quantifiers.
    
\end{itemize}

We say that a logic \(\cL\) is {\bf compact with inhabited sorts} in \(\cC\) if for every theory \(T\), \(T\) is consistent with inhabited sorts if and only if all finite subsets of \(T\) are consistent with inhabited sorts. The forward implication is immediate. We prove the converse by ultraproducts, as a consequence of \Los's theorem (Corollary~\ref{cor:Los theorem}).

\begin{Theorem}[Compactness by Ultraproducts]
Let \(\cC\) be a locally presentable two-valued topos whose terminal object \(\terminal\) is projective. Then coherent, regular, Cartesian, and positive logic are compact with inhabited sorts in \(\cC\). If \(\cC\) is further Boolean, then first-order logic is compact with inhabited sorts in \(\cC\).
\end{Theorem}

\begin{proof}
Let \(T\) be a theory such that every finite subset of \(T\) is consistent with inhabited sorts. Set
\[
    I = \powerset_{\mathrm{fin}}(T).
\]
For every \(i\in I\), choose an internally inhabited \(\Sigma\)-model
\(\cM_i\) of \(i\).

For every \(\varphi\in T\), let
\[
    I_\varphi = \set{J\in I\mid \varphi\in J}.
\]
The family \((I_\varphi)_{\varphi\in T}\) has the finite intersection property, since
\[
    \bigcap_{k=1}^n I_{\varphi_k} = I_{\set{\varphi_1,\ldots,\varphi_n}} \neq \emptyset.
\]
Hence, by the ultrafilter lemma, there exists an ultrafilter
\(F\) on \(I\) containing every \(I_\varphi\). Proposition~\ref{prop:filtered product is inhabited} ensures that the filtered product \(\prod_F \cM\) is internally inhabited.

Now, let \(\varphi\in T\). Since every \(J\in I_\varphi\) contains
\(\varphi\), we have
\[
    I_\varphi \subseteq \set{J\in I\mid \cM_J\models\varphi}.
\]
Since \(I_\varphi\in F\) and \(F\) is upward closed,
\[
    \set{J\in I\mid \cM_J\models\varphi}\in F.
\]
The appropriate case of Corollary~\ref{cor:Los theorem} gives \(\prod_F \cM \models \varphi\). Thus, the filtered product \(\prod_F \cM\) is an internally inhabited model of \(T\), and \(T\) is consistent with inhabited sorts in \(\cC\).
\end{proof}

\section{Conclusion}
\label{sec:conclusion}

In this paper, we explored ultraproducts of FOL structures categorically, extending \Los's theorem from set-theoretic models to Kripke--Joyal semantics and identified a framework in which the classical theorem extends to locally presentable topoi.
Local presentability of the ambient topos provides the interaction between filtered colimits and finite limits that the construction requires. At the same time, the internal inhabitedness of the structures and the projectivity of the terminal object allow the global handling of local witnesses.

Our result recovers the classical \Los's theorem in \(\Set\). At the same time, the examples of \(\Set\times\Set\) and \(G\text{-}\Set\) show that the hypotheses used in our proof are not necessary for \Los's property. Rather, they provide an internal mechanism to establish the closure properties needed for the proof. The detailed treatment of the logical connectives also illustrates the additional technical complexity that arises when classical model-theoretic arguments are expressed in the internal logic of a topos.
This technicality may also explain why classical model theory has seen limited generalizations (to the best of our knowledge) within categorical logic, beyond Gödel-style completeness theorems.

Our internal approach to reasoning about \Los's property suggests an external perspective on the problem of characterizing the topoi that have this property. Instead of constructing global witnesses internally, one may try to reason via logical points into \(\Set\), provided these points are sufficiently compatible with the ultraproduct construction.
Therefore, we conjecture that a locally presentable topos has \Los's property (for FOL) if and only if it admits a conservative family of logical points compatible with ultraproducts.

\section*{Acknowledgment}

During the preparation of the first version of this paper paper Romain Pascual was funded by the Deutsche Forschungsgemeinschaft (DFG, German Research Foundation) -- SFB 1608 - 501798263.

\printbibliography

@incollection{adamek_locally_1994,
	title        = {Locally {Presentable} {Categories}},
	author       = {Adamek, J. and Rosicky, J.},
	year         = 1994,
	booktitle    = {Locally {Presentable} and {Accessible} {Categories}},
	series       = {London {Mathematical} {Society} {Lecture} {Note} {Series}},
	pages        = {7--66},
	doi          = {10.1017/CBO9780511600579.004}
}

@article{AN78,
	title        = {L\`os Lemma Holds in Every Category},
	author       = {H. Andr\'eka and I. N\'emeti},
	year         = 1978,
	journal      = {Studia Scientiarum Mathematicarum Hungarica},
	volume       = 13,
	pages        = {361--376}
}

@article{barr_toposes_1974,
	title        = {Toposes without points},
	author       = {Barr, Michael},
	year         = 1974,
	journal      = {Journal of Pure and Applied Algebra},
	volume       = 5,
	number       = 3,
	pages        = {265--280},
	doi          = {10.1016/0022-4049(74)90037-1},
	issn         = {0022-4049}
}

@book{BW85,
	title        = {Toposes, triples and theories},
	author       = {M. Barr and C. Wells},
	year         = 1985
}

@book{BW90,
	title        = {Category Theory for Computing Science},
	author       = {M. Barr and C. Wells},
	year         = 1990,
	pages        = 325
}

@book{Car17,
	title        = {Theories, Sites, Toposes: Relating and studying mathematical theories through topos-theoretic 'bridges'},
	author       = {O. Caramello},
	year         = 2017
}

@book{Dia08,
	title        = {{I}nstitution-independent {M}odel {T}heory},
	author       = {R. Diaconescu},
	year         = 2008,
	series       = {Studies in Universal Logic},
	doi          = {10.1007/978-3-031-68854-6}
}

@article{Dia17,
	title        = {Implicit {Kripke} semantics and ultraproducts in stratified institutions},
	author       = {R. Diaconescu},
	year         = 2017,
	journal      = {Journal of Logic and Computation},
	volume       = 27,
	number       = 5,
	pages        = {1577--1606}
}

@book{johnstone_topos_1977,
	title        = {Topos Theory},
	author       = {Johnstone, P. T.},
	year         = 1977
}

@book{Johnstone02,
	title        = {Sketches of an Elephant: A Topos Theory Compendium. Vol.1 and Vol.2},
	author       = {P. Johnstone},
	year         = 2002
}

@article{keisler_ultraproduct_2010,
	title        = {The ultraproduct construction},
	author       = {H. J. Keisler},
	year         = 2009,
	booktitle    = {Ultrafilters across mathematics},
	series       = {Contemp. {Math}.},
	volume       = 530,
	pages        = {163--179},
	doi          = {10.1090/conm/530/10444}
}

@incollection{los_quelques_1955,
	title        = {Quelques remarques, th\'{e}or\`{e}mes et probl\`{e}mes sur les classes d\'{e}\-fi\-nis\-sables d'alg\`{e}bres},
	author       = {Jerzy Lo\'{s}},
	year         = 1955,
	booktitle    = {Studies in {Logic} and the {Foundations} of {Mathematics}},
	volume       = 16,
	pages        = {98--113},
	doi          = {10.1016/S0049-237X(09)70306-4}
}

@article{Mak87,
	title        = {Stone duality for first-order logic},
	author       = {M.Makkai},
	year         = 1987,
	journal      = {Advances in Math.},
	volume       = 65,
	pages        = {97--170},
	doi          = {10.1016/0001-8708(87)90020-X}
}

@book{McL71,
	title        = {Categories for the Working Mathematician},
	author       = {S. MacLane},
	year         = 1971
}

@book{mclarty_elementary_1992,
	title        = {Elementary Categories, Elementary Toposes},
	author       = {C. {McLarty}},
	year         = 1992,
	series       = {Oxford Logic Guides},
	doi          = {10.1093/oso/9780198533924.001.0001}
}

@book{MM12,
	title        = {Sheaves in geometry and logic: A first introduction to topos theory},
	author       = {S. MacLane and I. Moerdijk},
	year         = 2012
}

@article{Okh66,
	title        = {Ultrapowers in categories},
	author       = {T. Okhuma},
	year         = 1966,
	journal      = {Yokohama Mathematics Journal},
	volume       = 14,
	pages        = {17--37}
}

@article{sagi_ultraproducts_2023,
	title        = {Ultraproducts and {Related} {Constructions}},
	author       = {S\'{a}gi, G.},
	year         = 2023,
	journal      = {Mathematics},
	volume       = 11,
	number       = 1,
	pages        = 70,
	doi          = {10.3390/math11010070}
}

@article{Tar44,
	title        = {The {S}emantic {C}onception of {T}ruth},
	author       = {Tarski, A.},
	year         = 1944,
	journal      = {Philosophy and Phenomenological Research},
	volume       = 4,
	pages        = {13--47}
}

@article{Tar56,
	title        = {On the {C}oncept of {L}ogical {C}onsequence},
	author       = {Tarski, A.},
	year         = 1956,
	journal      = {Logic, Semantics, Metamathematics},
	pages        = {409--420},
	editor       = {Woodger, J.H.}
}

@book{verdier_theorie_1972,
	title        = {Th\'eorie Des {{Topos}} et {{Cohomologie Etale}} Des {{Sch\'emas}}. {{Tome}} 2. {{S\'eminaire}} de {{G\'eom\'etrie Alg\'ebrique}} Du {{Bois-Marie}} 1963-1964 ({{SGA}} 4)},
	author       = {Artin, M and Grothendieck, A. and Verdier, J. L.},
	year         = 1972,
	series       = {Lecture {{Notes}} in {{Mathematics}}},
	volume       = 270,
	doi          = {10.1007/BFb0061319}
}

\end{document}